\documentclass[a4paper,fleqn]{cas-dc}

\usepackage[numbers]{natbib}

\def\tsc#1{\csdef{#1}{\textsc{\lowercase{#1}}\xspace}}
\tsc{WGM}
\tsc{QE}
\tsc{EP}
\tsc{PMS}
\tsc{BEC}
\tsc{DE}

\newcommand{\argmax}{\mathop{\rm arg~max}\limits}

\begin{document}
\let\WriteBookmarks\relax
\def\floatpagepagefraction{1}
\def\textpagefraction{.001}
\shorttitle{BCI with RSMP using artificial facial images and voice}
\shortauthors{A Onishi et~al.}

\title [mode = title]{Brain-computer interface with rapid serial multimodal presentation using artificial facial images and voice}                      
\tnotemark[1]

\tnotetext[1]{This document is the results of a research
   project funded by the JSPS KAKENHI grants (18K17667, 21K11207).}

\author[1,2]{A Onishi}[type=editor,
                        auid=000,bioid=1,
                        prefix=Dr.,
                        role=Assistant Professor,
                        orcid=0000-0002-5566-1805]
\cormark[1]
\ead{onishi-a@es.kagawa-nct.ac.jp}
\ead[url]{http://onishi.starfree.jp/}

\credit{Conceptualization of this study, Methodology, 
Experiments, Data analysis, Writing}

\address[1]{Department of Electronic Systems Engineering, 
National Institute of Technology, Kagawa College, 
551, Kohda, Takuma-cho, Mitoyo-shi, Kagawa, 769-1192, Japan}
\address[2]{Center for Frontier Medical Engineering, Chiba University, 
1-33 Yayoi-cho, Inage-ku, Chiba, Japan}

\cortext[cor1]{Corresponding author}

\begin{abstract}
Electroencephalography (EEG) signals elicited by multimodal stimuli 
can drive brain-computer interfaces (BCIs), and research has demonstrated 
that visual and auditory stimuli can be employed simultaneously to improve BCI performance. 
However, no studies have investigated the effect of multimodal stimuli 
in rapid serial visual presentation (RSVP) BCIs. 
In the present study, we propose a rapid serial multimodal presentation (RSMP) BCI 
that incorporates artificial facial images and artificial voice stimuli. 
To clarify the effect of audiovisual stimuli on the RSMP BCI, 
scrambled images and masked sounds were applied 
instead of visual and auditory stimuli, respectively. 
Our findings indicated that the audiovisual stimuli improved the performance of the RSMP BCI, 
and that the P300 at Pz contributed to classification accuracy. 
Online accuracy of BCI reached 85.7$\pm$11.5\%. 
Taken together, these findings may aid in the development of better gaze-independent BCI systems.
\end{abstract}

\begin{highlights}
\item We proposed a P300-based RSMP BCI that uses artificial face and voice stimuli. 
\item Audiovisual stimuli enhanced the classification accuracy for the RSMP BCI. 
\item P300 at Pz contributed to the classification of the BCI. 
\end{highlights}

\begin{keywords}
BCI \sep P300 \sep RSMP \sep RSVP \sep multimodal \sep audiovisual
\end{keywords}

\maketitle

\section{Introduction}
\noindent
Brain-computer interfaces (BCIs) measure brain signals, which are then decoded into commands for controlling an external device \cite{Wolpaw2002}, making them valuable for individuals with disabilities. 
Several BCIs that rely on electroencephalography (EEG) have been proposed. 
One well-studied BCI utilizes the P300 component of the event-related potential (ERP)(i.e., P300- or ERP-based BCI), which appears in response to rare stimuli \cite{farwell1988talking}. 

P300-based BCIs can be driven by visual, auditory, or tactile stimuli. 
In early studies, visual stimuli included a $6\times6$ matrix of letters, 
which is referred to as the P300 speller or Farwell and Donchin speller \cite{farwell1988talking, schalk2004bci2000, Salvaris2009}. 
The P300 speller turns a row or column of gray letters on the matrix white. 
Users can spell a desired letter by counting silently when the letter turns white. 
Similarly, an auditory P300-based BCI that can select 
``Yes'', ``No'', ``Pass'', and ``End'' has been investigated \cite{Sellers2006}. 
Furthermore, tactors attached to a participant's waist can also 
be used instead of visual or auditory stimuli \cite{Brouwer2010}. 
These three sensory modalities are associated with different pathways to the brain. 
Thus, even with impairment in one modality, P300-based BCIs can be effective. 

Performance of the BCI depends on the stimulus 
because the ERP is modulated by stimulus modality and content, and because the ERP including the P300 is used as a feature for the classification of the BCI. 
P300 amplitude and latency differ across modalities, 
exhibiting differences in classification accuracy \cite{belitski2011p300, Thurlings2012}.
In addition, complex visual and auditory stimuli 
that contain rich information have been applied to BCI. 
In one previous study, a P300-based BCI with a green/blue flicker matrix exhibited a higher accuracy than one with a white/gray flicker matrix \cite{Takano2009}. 
Another study reported improved performance using a P300-based BCI that presents facial images instead of color changes \cite{Kaufmann2011, Onishi2011, Jin2012}. Several studies have also investigated the applicability of auditory stimuli. 
Spatial auditory stimuli from speakers around a user are helpful for increasing the accuracy of P300-based BCIs \cite{Schreuder2010}. 
Furthermore, natural auditory stimuli, such as animal sounds (e.g., frog, seagull), exhibit unique ERP waveforms \cite{simon2015auditory}. 
These sensory modalities can also be applied to the BCI simultaneously. 

Research has demonstrated that visual and auditory stimuli can be employed simultaneously in BCIs \cite{Sellers2006}. For example, an audiovisual BCI that responds yes or no has been proposed \cite{Wang2015}. Another study indicated that a bimodal P300-based BCI combining visual and tactile stimuli exhibited better performance than unimodal BCIs \cite{Brouwer2010, Thurlings2012}. 
Furthermore, auditory stimuli delivered via bone conduction headphones 
and tactile stimuli have been applied in multimodal BCI systems \cite{Rutkowski2015}. 
A bimodal, direction-congruent BCI with spatial auditory stimuli and corresponding tactile stimuli 
exhibited better performance than unimodal BCIs \cite{Yin2016}. 
Taken together, these findings indicated that multimodal stimuli improve the classification accuracy of the BCI. Thus, multimodal BCIs are advantageous in that they can not only use multiple sensory pathways 
but also improve performance, likely via sensory integration. 

Unlike the letter matrices utilized in traditional visual P300-based BCIs, rapid serial visual presentation (RSVP) involves the rapid presentation of stimuli at the center of 
the monitor one by one in a random order \cite{Acqualagna2010}.  
RSVP is advantageous for P300-based systems
because it does not require eye gaze movements to drive the BCI \cite{Thurlings2012}. 
However, the effect of multimodal stimuli on the performance of the BCI remains unclear. 

Therefore, in the present study, we proposed a P300-based BCI incorporating 
rapid serial multimodal presentation (RSMP) (RSMP BCI), 
which utilizes artificial facial images and artificial voice. 
The stimuli represented five Japanese vowels, and they were provided 
such that each stimulus indicated a single vowel.
We hypothesized that audiovisual stimuli 
would also be effective for the P300-based RSMP BCI given the congruence of the stimuli 
and ERP components elicited by the facial images.
To clarify the effect of audiovisual integration in BCI systems, 
we prepared and compared facial images with phase scrambling as well as masked sounds. 
Furthermore, the online classification accuracy and the effect of the stimulus onset asynchrony were evaluated.

\section{Methods}
\noindent
This study comprised of two experiments.
Experiment 1 intended to reveal the effect of stimulus types. 
On the other hand, experiment 2 aimed to reveal the accuracy of online classification 
and the effect of stimulus onset asynchrony.

\subsection{Experiment 1: Comparison of stimulus types}

\subsubsection{Participants}
\noindent
Eleven healthy participants were included in experiment 1 (exp.~1). 
Their mean age was $24.0\pm 3.0$ years. 
Two of them were female, while the others were male. 
Two participants were left-handed. 
They all had normal or corrected visual acuity in addition to auditory acuity. 
All participants provided written informed consent prior to the experiment. 
This experiment took place at Chiba University. 
This study was conducted in accordance with the guidelines of the Internal Ethics Committee at Chiba University. 
 
\subsubsection{Stimuli}
\begin{figure}
	\centering
		\includegraphics[scale=.2]{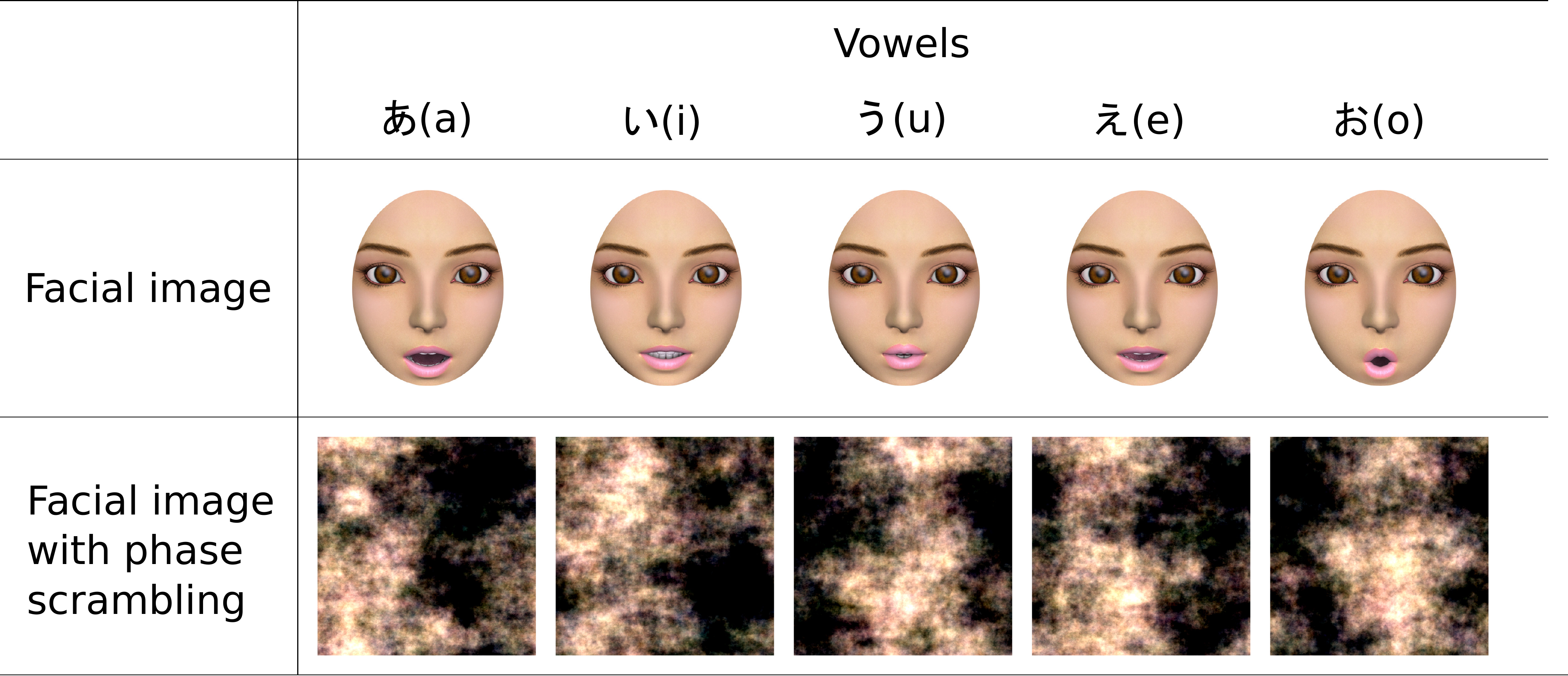}
	\caption{Types of visual stimuli. Artificial facial images that 
	represent Japanese vowels were rendered, following which they were trimmed in an ellipse.
	These stimuli were used in the AV and V conditions. 
	In addition, phase scrambling was applied to the facial images used in the A condition.  
	}
	\label{fig:type}
\end{figure}

\noindent
In this study, we prepared an RSMP BCI system that can select one of 
five Japanese vowels using brain signals. 
Artificial facial images and vowel sounds were applied as BCI stimuli. We examined the effect of the audiovisual stimuli on the RSMP BCI by comparing the following three conditions: audiovisual (AV), visual (V), and auditory (A). 
In the AV condition, the artificial facial images shown in Fig.~\ref{fig:type} 
and corresponding artificial voice were presented simultaneously. 
In the V condition, artificial voice stimuli were masked and presented with the artificial facial images. 
In the A condition, facial images with phase scrambling (see Fig.~\ref{fig:type}) 
were presented together with the artificial voice. 
Note that both visual and auditory stimuli were presented simultaneously 
even in the V and A conditions. 

\begin{figure}
	\centering
		\includegraphics[scale=.2]{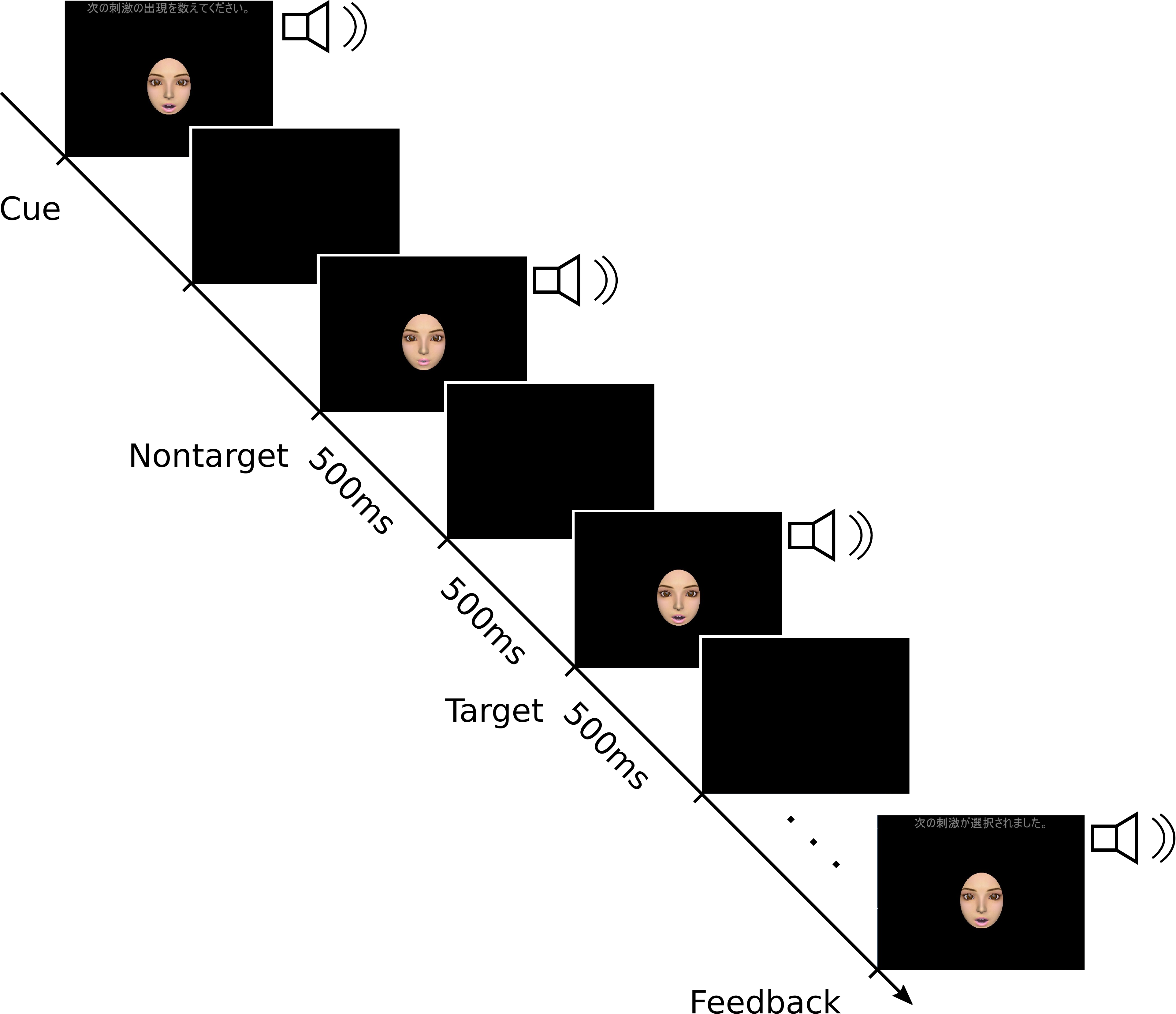}
	\caption{Example of the BCI task (AV condition). 
	First, a vowel to count (target vowel) was cued with audiovisual stimuli and an instruction message. 
	Second, a facial image and the corresponding sound were presented for 500 ms, 
	following which they disappeared for 500 ms. 
	Such stimuli appeared every 1,000 ms, changing vowels in a pseudo-random order.  
	If the stimuli represented the cued vowel, 
	a participant counted the appearance of the stimuli silently (up to 15 times). 
	After stimulus presentation, EEG signals recorded during the above task were 
	translated into a vowel and fed back to the participant if the classifier was trained. 
	}
	\label{fig:task}
\end{figure}

To input a cued Japanese vowel via this BCI system, 
a participant was asked to count the appearance of 
the instructed vowel in response to series of stimuli. 
Figure \ref{fig:task} represents an example of the task. 
At the beginning, a cue was presented to the participant. 
Next, audiovisual stimuli were presented in random order. 
During stimulus presentation, participants counted the appearances of the 
cued stimuli silently. 
Finally, EEG signals recorded during the task were analyzed and translated into a vowel output. 
The output was fed back to participants online only in exp.~2.  
Note that all participants were informed regarding the cue type, which could be recognized from the auditory or visual stimulus in the A and V conditions. 

This experiment was conducted in a soundproof room. 
Participants were seated on a comfortable chair located 60 cm from the monitor. 
Each stimulus lasted 500 ms, and the inter-stimulus interval was 500 ms. 
In other words, the stimulus onset asynchrony (SOA) was 1,000 ms. 
All five stimuli were repeatedly presented 15 times in a trial. 
Each run included two trials.
Runs were repeated five times in a pseudo-random order, 
respectively for each stimulus condition.  
The probability of target appearance was 1/5. 
During the experiment, the output was not indicated in order to save experimental time and to fatigue among participants. 
Classification accuracy for each stimulus condition was 
calculated via offline cross-validation. 
Before starting BCI experiments, 
the sound pressure level was adjusted to 20 dB SL 
by the Method of Limits. 

The BCI system consisted of a
PC for managing the experiment and EEG recordings (HP Probook 430 G3, HP Inc., CA), a PC for presenting the stimulus (Handmade PC, G31M-ES2L, Windows 10), a display monitor (E178FPc, Dell computers, TX), an EEG amplifier (Polymate mini AP108, Miyuki-Giken Co., Ltd., Japan), 
an audio interface (UCA222, Behringer GmbH, Germany), headphones (ATH-M20x, Audio-Technica Co., Ltd., Japan), 
and an AD-converter (AIO-160802AY-USB, Contec Co., Ltd., Japan). 
The sound pressure level was adjusted using an attenuator (FX-AUDIO AT-01J, North Flat Japan Co., Ltd., Japan). 
The BMI system was coded with C and Python with Psychopy. 
The above-mentioned audiovisual stimuli were generated as follows:
For facial images, 
a three-dimensional (3D) human model TY2 was rendered using Poser 10 (Smith Micro Inc., CA). 
Then, the facial images were trimmed elliptically using Corel PHOTO-PAINT Essentials X8 (Corel Inc., Canada). 
For the contrast experiment, the images were scrambled via 
Fourier phase scrambling using MATLAB 2016a (MathWorks Inc., MA). 
Artificial voice stimuli were generated with
CeVIO Creative Studio S 6 (Frontier works Inc., Japan)
using the ONE model. 
All vowels were C4 in musical notes and lasted for 500 ms. 
The sounds were trimmed, and the RMSs were equalized using MATLAB and employed in AV and A conditions. 
Sounds masked by Gaussian noise were also generated in MATLAB for use in the V condition.

\subsection{Experiment 2: Effect of stimulus onset asynchrony}

\subsubsection{Participants}
\noindent
Seven healthy participants were included in experiment 2 (exp.~2), where one of them also participated in exp.~1. 
Their mean age was $21.1\pm 5.3$ years. 
Two females participated in the experiment. 
One participant was left-handed. 
Their vision was all normal or corrected, and their hearing was normal. 
All participants provided written informed consent prior to the experiment. 
This experiment was conducted at the National Institute of Technology, Kagawa College.
This study was conducted in accordance with the guidelines of 
the Ethics Committee at Shikoku Kosen Center for Innovative Technologies.

\subsubsection{Stimuli}
\noindent
To investigate the effect of SOA, and to measure the online classification accuracy,  
three conditions of SOA were prepared; 
1000, 250, and 100 ms. 
In exp.~2, the audiovisual stimuli used in exp.~1 were presented.
Note, the ratio of the stimulus duration and the ISI was kept (1:1). 

Most of the stimulus parameters used in this experiment were the same as in exp.~1. 
The difference of parameters or conditions between exp.~1 and 2 were as follows: 
Each run included three trials. 
The training session contained 5 runs, while the testing session included 2 runs. 
In the testing session, outputs estimated from EEG were fed back to each participant.  
This experiment was conducted outside the soundproof room. 
Stimulus was presented by a display monitor (LCD-AD173SESW, I-O DATA Co., Ltd, Japan),

\subsection{EEG recording}
\noindent
EEG signals were recorded from C3, Cz, C4, P3, Pz, P4, O1, and O2, 
where the ground and reference electrodes were the forehead and right mastoid, respectively. 
Active electrodes were used for measuring EEG. 
The impedance was kept below 10 k$\Omega$. 
The sampling rate was 500 Hz. 
A hardware low-pass filter (cut-off frequency: 30 Hz) and 
high-pass filter (time constant: 1.5 sec.) were applied in addition to a notch-filter (50 Hz for exp.~1 and 60 Hz for exp.~2).

\subsection{Offline classification accuracy}
\noindent
Offline classification accuracy of exp.~1 was estimated via offline 
leave-one-out cross-validation using entire data of exp.~1, while separated training and testing data recorded in exp.~2 were used for computing offline accuracy of exp.~2. 
Recorded EEG signals were trimmed for 1 s, and the baseline correlation was removed using pre-0.1 s EEG signals. 
Before classification, a notch filter (50 Hz for exp.~1, 60 Hz for exp.~2), Savitzky-Golay filter (3rd order, 69 sample), and downsampling (140 samples) were applied. 
The multichannel EEG data were then vectorized. 
Principal component analysis was applied to reduce 
the number of dimensions, where the contribution threshold was 0.9999. 
Finally, linear discriminant analysis (LDA) was applied to classify the signal.

The output was determined as follows.
The vowels ``a,'' ``i,'' ``u,'' ``e,'' and ``o,'' were labeled as 1 to 5, respectively. 
Given the stimulus set $I\in \left\lbrace 1, 2, ..., 5\right\rbrace$, 
the number of stimulus repetitions $R$, 
preprocessed and vectorized testing data in response to stimulus $i$ of the $r$-th stimulus repetition $\bold{x}_{r,i}$, 
and the trained weight vector of the LDA $\bold{w}$, 
the input was estimated by finding the stimulus $i$ yielding the maximum summation of the inner product of data and LDA weight vector: 

\begin{eqnarray}
\hat{i}=\argmax_{i \in I} \sum_{r=1}^{R} \mathbf{w} \cdot \mathbf{x}_{r, i}. 
\end{eqnarray}

\noindent
The estimate stimuli $\hat{i}$ was the output of the trial. 
For example, the vowel ``u'' was the output if $\hat{i}=3$. 
If the output was equal to a cued stimulus, the estimation was correct. 
The classification accuracy was decided by $\#correct/\#output$. 
During the offline analysis, the number of stimulus repetitions $R$ was varied from 1 to 15 
to estimate accuracy for each stimulus repetition. 
The offline accuracy was statistically analyzed using two-way repeated-measures analyses of variance (ANOVAs). 
Note that the factors included were stimulus condition and the number of stimulus repetitions. 
In addition, post-hoc pair-wise t-tests were applied, where p-values were corrected using Bonferroni's method.

\subsection{Online classification accuracy}
\noindent
The method used in the offline analysis was also applied in the online classification of exp.~2. 
The online classification accuracy was also decided by $\#correct/\#output$. 
Note that $R$ was fixed to 15. 
The online classification accuracy was statistically analyzed by the one-way repeated-measures ANOVAs. 
Moreover, post-hoc pair-wise t-test was used in which Bonferroni's method was applied.

\subsection{Information transfer rate}
\noindent
Information transfer rate (ITR) indicates the amount of information that is transferred, 
which is frequently used as a performance indicator of the BCIs \cite{Wolpaw2002,zhang2008asynchronous}. 
The ITR took account classification accuracy, time for output, and the number of output commands. 
Given time for cue $T_{cue} = 2$ [s], stimulus onset asynchrony $T_{SOA}$ [s], 
and ERP buffer length $T_{buffer}=1$ [s], the number of stimuli (the number of output command) $N_s=n(I)=5$, 
period to output a command by a P300 BCI is calculated by $T=T_{cue}+(T_{SOA})\times(R \times N_s-1)+T_{buffer}$ [s]. 
In other words, the number of output per minute is $M=\frac{60}{T}$. 
The ITR $B$ [bit/min] is computed by 
\begin{equation}
B=M \left\lbrace \log_2 N_s + A \log_2 A + (1-A) \log _2 \left( \frac{1-A}{N_s-1} \right) \right\rbrace, 
\end{equation}
where $A$ denotes classification accuracy. 
The ITR increases as the $A$ and $N_s$ increase, and as $T$ or $R$ decreases. 
Statistical analysis was also applied to the ITR as well as offline and online accuracy.

\subsection{EEG waveform analysis}
\noindent
To gain insight into the contribution of EEG waveforms, 
grand-averaged EEG waveforms were visualized. 
Similar to the offline classification, 
baseline correction, a notch filter, and a Savitzky-Golay filter were applied. 
Note that downsampling and PCA were not applied when computing grand-averaged EEG waveforms. 

In addition, the point-biserial correlation coefficients or $r^2$-values were computed \cite{blankertz2011single, Onishi2017, Tate1954}. 
The $r^2$-value of a time sample in a channel can be calculated as follows: 
Given the amount of data in target and nontarget classes $N_2$ and $N_1$, 
mean values of target and nontarget classes $\mu_2$ and $\mu_1$, 
and standard deviation $\sigma$, 
the $r$-value is computed as
 
\begin{eqnarray}
r:= \frac{\sqrt{N_2 \cdot N_1}}{N_2 + N_1} \frac{\mu_2 - \mu_1}{\sigma}. 
\end{eqnarray}

\noindent
The $r$-value stands for the Pearson correlation between ERP amplitude and classes. 
This implies that the statistical test for the Pearson correlation is applicable. 
We applied a test of no correlation, where p-values were corrected using Bonferroni's method. 
The $r$-value was squared ($r^2$-value) to achieve visualization. 
The $r^2$-value increases as the mean values of the target and nontarget classes separate, and as the standard deviation decreases. 
Note that downsampling was applied when computing $r^2$-values in addition to the 
preprocessing applied to the grand-averaged EEG waveforms, 
but PCA was not applied.

\section{Results}

\subsection{Experiment 1}

\subsubsection{Offline classification accuracy} 
\begin{figure}
	\centering
		\includegraphics[scale=.5]{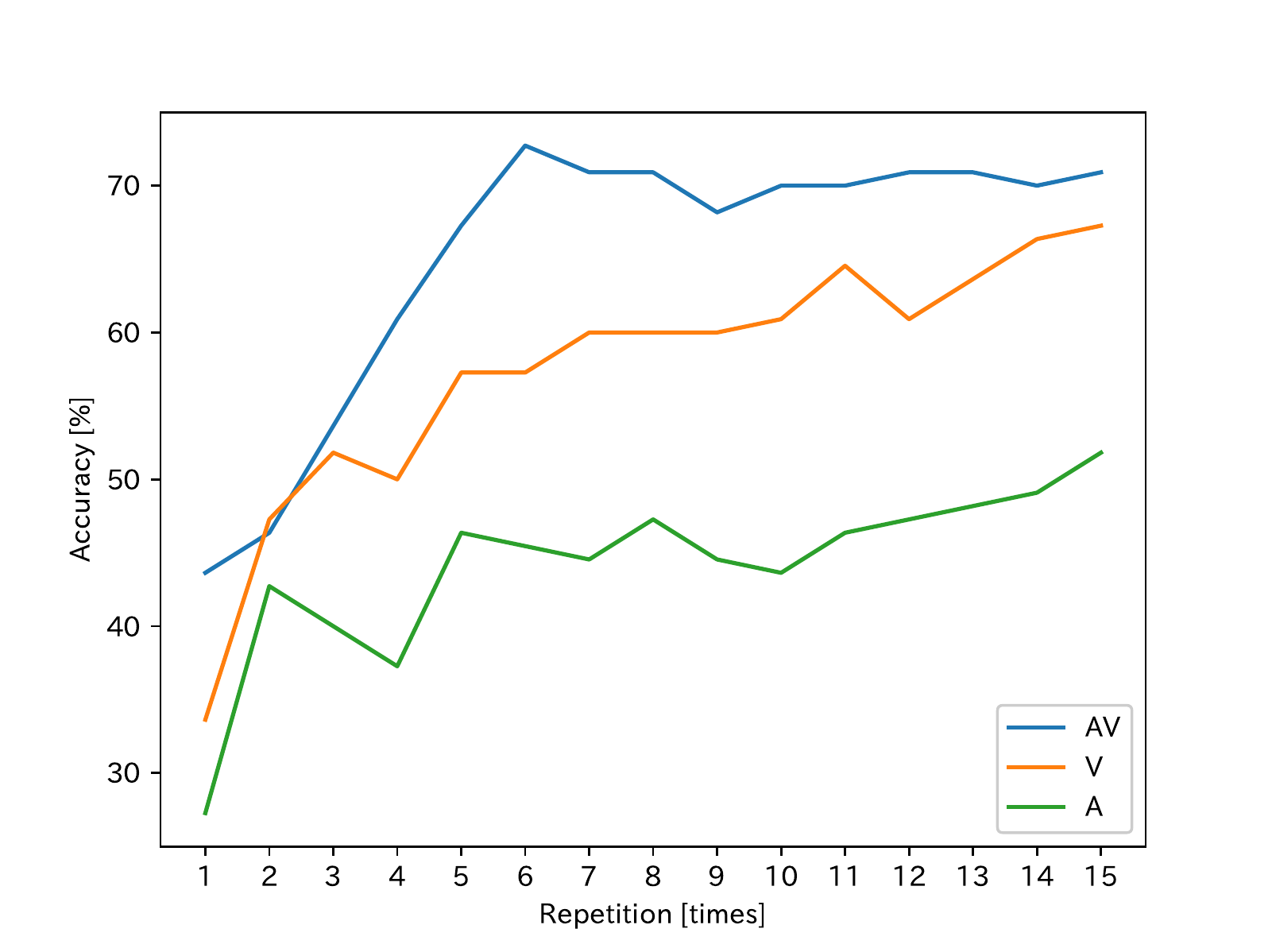}
	\caption{Offline mean classification accuracy for audiovisual (AV), visual (V), and auditory (A)
	conditions.}
	\label{FIG:Exp1_Accuracy}
\end{figure}

\noindent
In order to clarify the effect of audiovisual integration on the RSMP BCI, 
we calculated and compared classification accuracy among the AV, V, and A conditions.
The offline classification accuracy of these three conditions is shown in Fig.~\ref{FIG:Exp1_Accuracy} and Table.~\ref{Table:Exp1_acccuracy}. 
The highest mean accuracy was observed in the AV condition, followed by the V condition in which stimuli were presented more than three times. The lowest mean accuracy for all repetitions was observed in the A condition. 
AV, V, and A conditions reached 72.7\%, 67.3\%, and 51.8\% at best. 

Two-way repeated measures ANOVA revealed significant main effects 
of stimulus type ($p<0.05$, $F(2,20)$ = 5.259) and repetition ($p<0.001$, $F(14,140)$ = 12.78). 
The post-hoc pairwise t-test revealed significant differences between all pairs of stimulus type ($p<0.001$).

\begin{table*}[width=2.0\linewidth,cols=17,pos=h]
\caption{Offline classification accuracy and standard deviation (SD) for each stimulus condition and the number of repetition.}
\label{Table:Exp1_acccuracy}
\begin{tabular*}{\tblwidth}{@{} CC|CCCCCCCCCCCCCCC@{} }
\toprule
Condition & Subject & \multicolumn{15}{c}{Repetition} \\
~ & ~ & 1 & 2 & 3 & 4 & 5 & 6 & 7 & 8 & 9 & 10 & 11 & 12 & 13 & 14 & 15 \\
\hline
AV 	& 1 	& 30	& 20	& 20	& 50	& 50	& 70	& 60	& 80	& 50	& 60	& 60	& 60	& 60	& 40	& 40 \\
~	& 2 	& 20	& 50	& 50	& 50	& 50	& 60	& 40	& 40	& 50	& 40	& 50	& 50	& 50	& 50	& 50 \\
~	& 3 	& 50	& 50	& 40	& 70	& 70	& 60	& 80	& 60	& 60	& 70	& 70	& 80	& 80	& 70	& 80 \\
~	& 4 	& 30	& 50	& 50	& 50	& 80	& 90	& 80	& 90	& 90	& 90	& 90	& 100	& 100	& 100	& 100 \\
~	& 5 	& 60	& 70	& 90	& 90	& 90	& 90	& 90	& 90	& 90	& 90	& 90	& 90	& 90	& 90	& 90 \\
~	& 6 	& 20	& 30	& 40	& 40	& 60	& 60	& 70	& 50	& 40	& 50	& 50	& 20	& 30	& 30	& 30 \\
~	& 7 	& 20	& 10	& 20	& 20	& 20	& 50	& 40	& 40	& 30	& 40	& 30	& 40	& 30	& 40	& 30 \\
~	& 8 	& 70	& 90	& 80	& 80	& 80	& 80	& 80	& 70	& 70	& 70	& 70	& 80	& 80	& 80	& 90 \\
~	& 9 	& 70	& 50	& 80	& 90	& 100	& 90	& 90	& 90	& 100	& 100	& 100	& 100	& 100	& 100	& 100 \\
~	& 10 	& 50	& 30	& 50	& 70	& 70	& 70	& 80	& 90	& 90	& 80	& 80	& 80	& 80	& 80	& 80 \\
~	& 11	& 60	& 60	& 70	& 60	& 70	& 80	& 70	& 80	& 80	& 80	& 80	& 80	& 80	& 90	& 90 \\
~	& Mean	& 43.6	& 46.4	& 53.6	& 60.9	& 67.3	& 72.7	& 70.9	& 70.9	& 68.2	& 70.0	& 70.0	& 70.9	& 70.9	& 70.0	& 70.9 \\
~	& SD	& 20.1	& 22.9	& 23.8	& 21.7	& 22.0	& 14.2	& 17.6	& 20.2	& 23.6	& 20.5	& 21.0	& 25.5	& 25.1	& 25.7	& 27.7 \\
\hline
V 	& 1 	& 30	& 20	& 10	& 10	& 0		& 10	& 20	& 0		& 0		& 10	& 20	& 10	& 30	& 30	& 30 \\
~ 	& 2 	& 10	& 20	& 10	& 30	& 20	& 20	& 40	& 40	& 40	& 50	& 60	& 30	& 40	& 50	& 50 \\
~ 	& 3 	& 50	& 80	& 70	& 60	& 80	& 80	& 80	& 80	& 80	& 80	& 90	& 80	& 90	& 90	& 100 \\
~ 	& 4 	& 20	& 40	& 60	& 80	& 80	& 80	& 90	& 100	& 100	& 100	& 100	& 100	& 100	& 100	& 100 \\
~ 	& 5 	& 30	& 50	& 70	& 70	& 80	& 60	& 70	& 80	& 70	& 60	& 60	& 70	& 80	& 80	& 80 \\
~ 	& 6 	& 20	& 40	& 50	& 30	& 50	& 50	& 40	& 40	& 40	& 30	& 50	& 50	& 40	& 60	& 60 \\
~ 	& 7 	& 40	& 30	& 20	& 30	& 30	& 40	& 30	& 30	& 30	& 50	& 30	& 40	& 40	& 40	& 40 \\
~ 	& 8 	& 30	& 70	& 70	& 60	& 80	& 80	& 70	& 80	& 80	& 80	& 90	& 90	& 90	& 90	& 90 \\
~ 	& 9 	& 60	& 50	& 60	& 60	& 60	& 60	& 60	& 50	& 60	& 50	& 50	& 40	& 40	& 40	& 50 \\
~ 	& 10 	& 20	& 30	& 50	& 50	& 70	& 60	& 60	& 60	& 60	& 60	& 60	& 60	& 50	& 50	& 40 \\
~ 	& 11 	& 60	& 90	& 100	& 70	& 80	& 90	& 100	& 100	& 100	& 100	& 100	& 100	& 100	& 100	& 100 \\
~	& Mean	& 33.6	& 47.3	& 51.8	& 50.0	& 57.3	& 57.3	& 60.0	& 60.0	& 60.0	& 60.9	& 64.5	& 60.9	& 63.6	& 66.4	& 67.3 \\
~	& SD	& 16.9	& 23.7	& 28.2	& 21.9	& 28.7	& 25.7	& 25.3	& 31.3	& 30.7	& 27.7	& 27.3	& 29.8	& 28.0	& 26.2	& 27.2 \\
\hline
A 	& 1 	& 30	& 30	& 20	& 10	& 30	& 20	& 10	& 20	& 20	& 10	& 10	& 0		& 0		& 0		& 0 \\
~ 	& 2 	& 30	& 40	& 20	& 60	& 30	& 30	& 20	& 20	& 20	& 20	& 30	& 20	& 30	& 40	& 40 \\
~ 	& 3 	& 10	& 40	& 20	& 10	& 10	& 10	& 10	& 20	& 20	& 20	& 30	& 20	& 20	& 30	& 30 \\
~ 	& 4 	& 20	& 10	& 40	& 20	& 30	& 30	& 30	& 30	& 20	& 10	& 30	& 50	& 30	& 20	& 40 \\
~ 	& 5 	& 40	& 70	& 60	& 40	& 60	& 60	& 60	& 50	& 60	& 50	& 50	& 40	& 60	& 70	& 60 \\
~ 	& 6 	& 30	& 50	& 40	& 40	& 50	& 60	& 50	& 60	& 40	& 40	& 40	& 50	& 40	& 40	& 50 \\
~ 	& 7 	& 10	& 50	& 30	& 30	& 30	& 30	& 50	& 60	& 50	& 60	& 50	& 50	& 60	& 60	& 70 \\
~ 	& 8 	& 20	& 50	& 70	& 40	& 50	& 50	& 60	& 50	& 50	& 60	& 60	& 70	& 70	& 60	& 60 \\
~ 	& 9 	& 50	& 40	& 30	& 40	& 50	& 40	& 50	& 50	& 60	& 60	& 70	& 80	& 70	& 70	& 70 \\
~ 	& 10 	& 40	& 30	& 40	& 50	& 70	& 70	& 50	& 60	& 50	& 50	& 40	& 40	& 50	& 50	& 50 \\
~ 	& 11 	& 20	& 60	& 70	& 70	& 100	& 100	& 100	& 100	& 100	& 100	& 100	& 100	& 100	& 100	& 100 \\
~	& Mean	& 27.3	& 42.7	& 40.0	& 37.3	& 46.4	& 45.5	& 44.5	& 47.3	& 44.5	& 43.6	& 46.4	& 47.3	& 48.2	& 49.1	& 51.8 \\
~	& SD	& 12.7	& 16.2	& 19.0	& 19.0	& 24.6	& 25.8	& 26.2	& 24.1	& 24.6	& 27.3	& 24.2	& 28.7	& 27.9	& 27.4	& 25.6 \\
\bottomrule
\end{tabular*}
\end{table*}

\subsubsection{Offline ITR} 
\begin{figure}
	\centering
		\includegraphics[scale=.5]{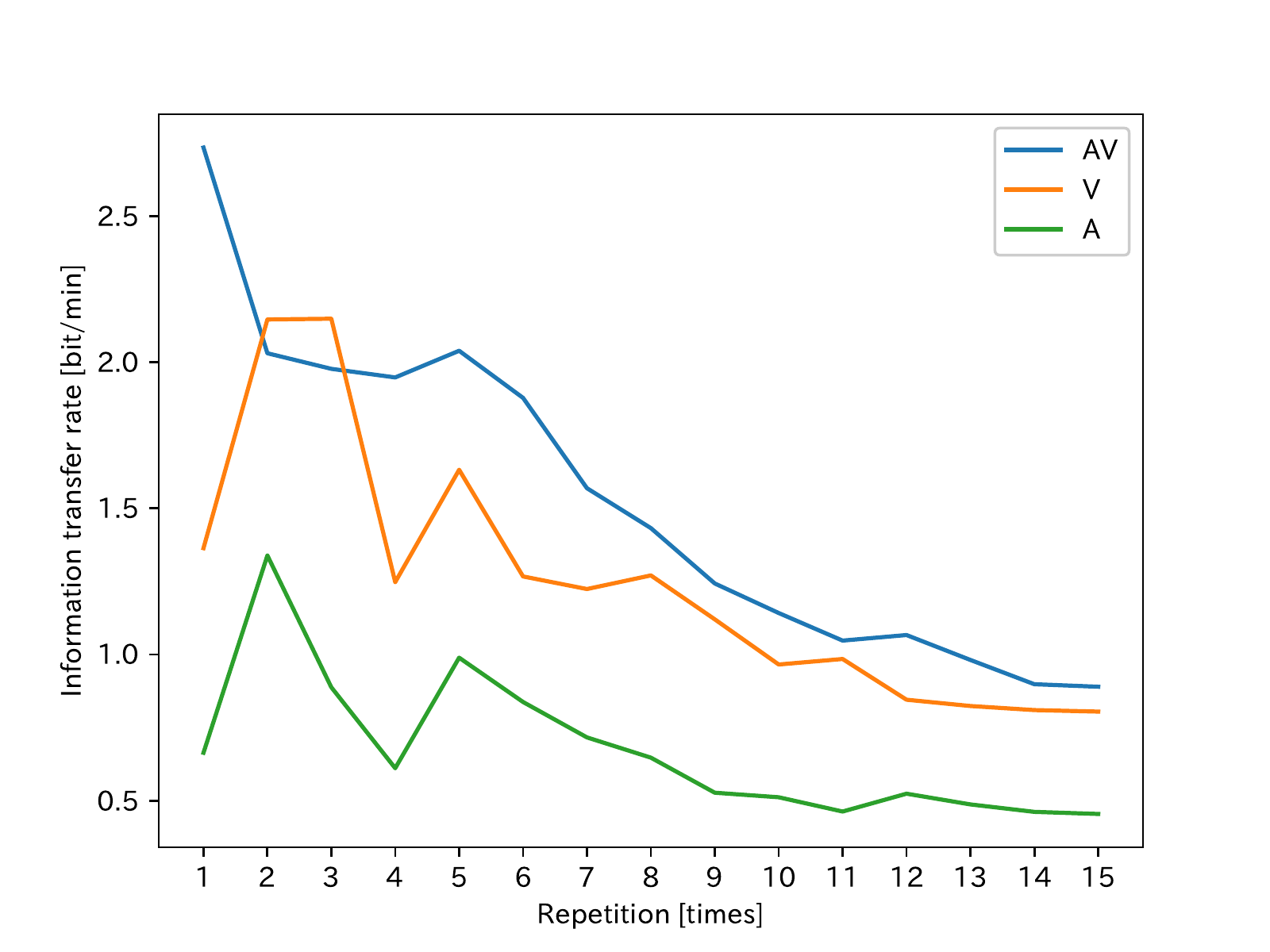}
	\caption{Offline mean ITR for audiovisual (AV), visual (V), and auditory (A)
	conditions.}
	\label{FIG:Exp1_ITR}
\end{figure}

\noindent
The mean offline ITR of exp.~1 is shown in Fig.~\ref{FIG:Exp1_ITR}. 
AV condition showed the highest mean ITR of all conditions. 
V condition showed the second, and A condition represented the worst ITR.  
AV condition with repetition 1 indicated the highest mean ITR (2.73 bit/min).

Two-way repeated measures ANOVA revealed significant main effects 
of stimulus type ($p<0.05$, $F(2,20)$ = 4.514), repetition ($p<0.001$, $F(14,140)$ = 5.612), and their interaction ($p<0.05$, $F(28,280)$ = 1.668). 
The post-hoc pairwise t-test revealed significant differences between all pairs of stimulus type ($p<0.05$). 

\subsubsection{Grand averaged EEG waveforms}
\begin{figure}
	\centering
		\includegraphics[scale=.5]{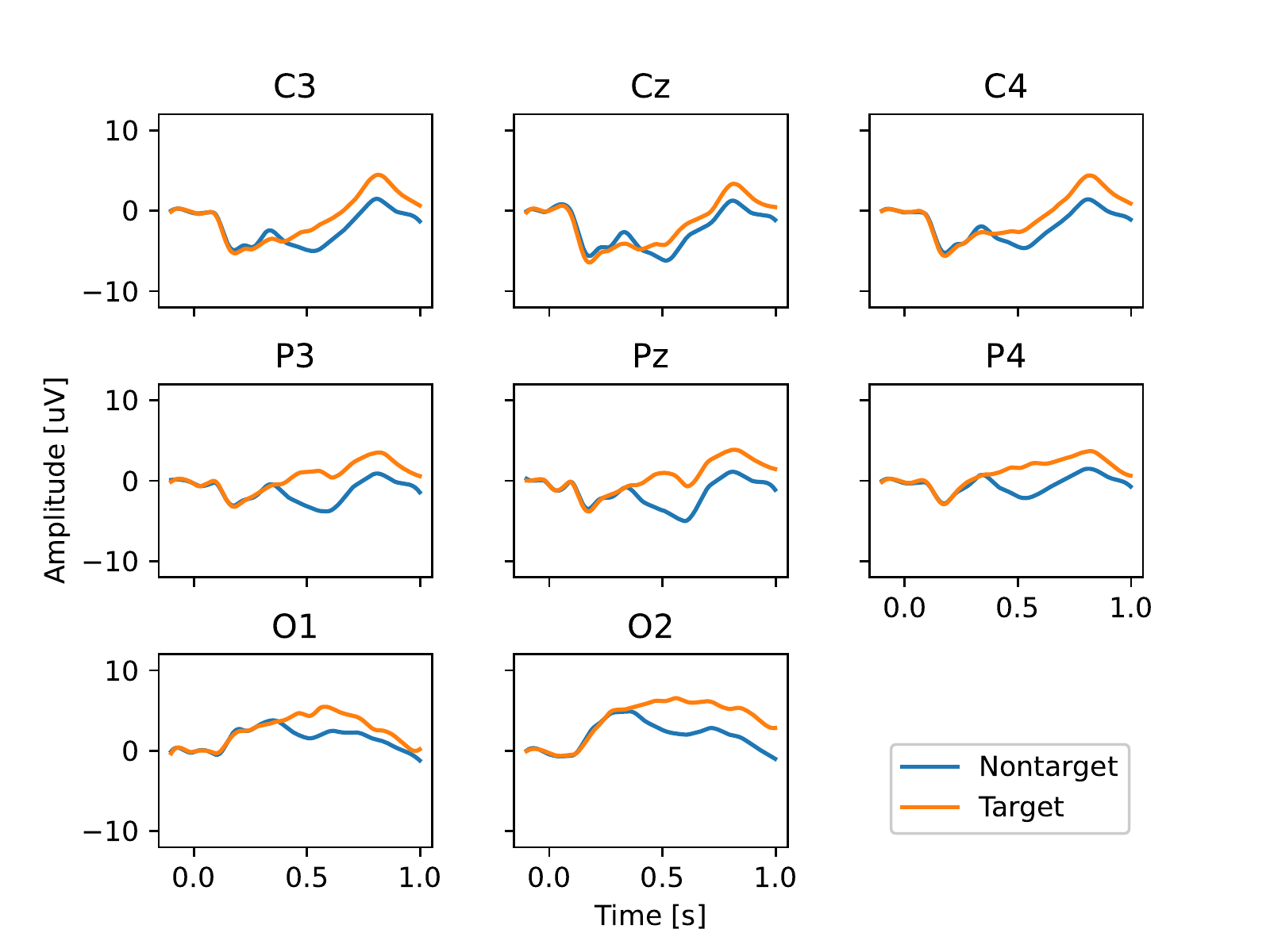}
	\caption{Grand-averaged EEG waveforms in the AV condition. }
	\label{FIG:Exp1_GrandAverageAV}
\end{figure}

\begin{figure}
	\centering
		\includegraphics[scale=.5]{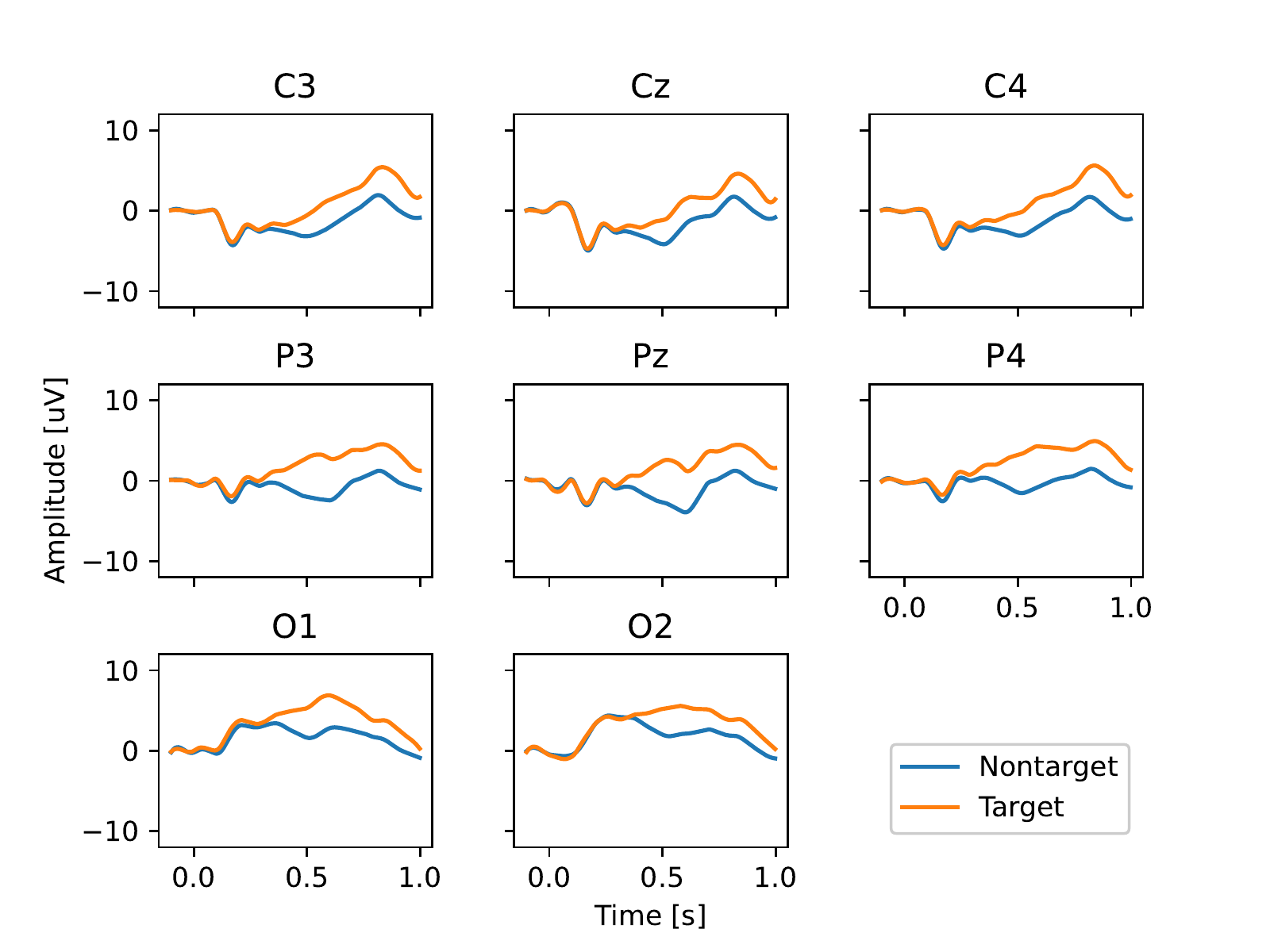}
	\caption{Grand-averaged EEG waveforms in the V condition. }
	\label{FIG:Exp1_GrandAverageV}
\end{figure}

\begin{figure}
	\centering
		\includegraphics[scale=.5]{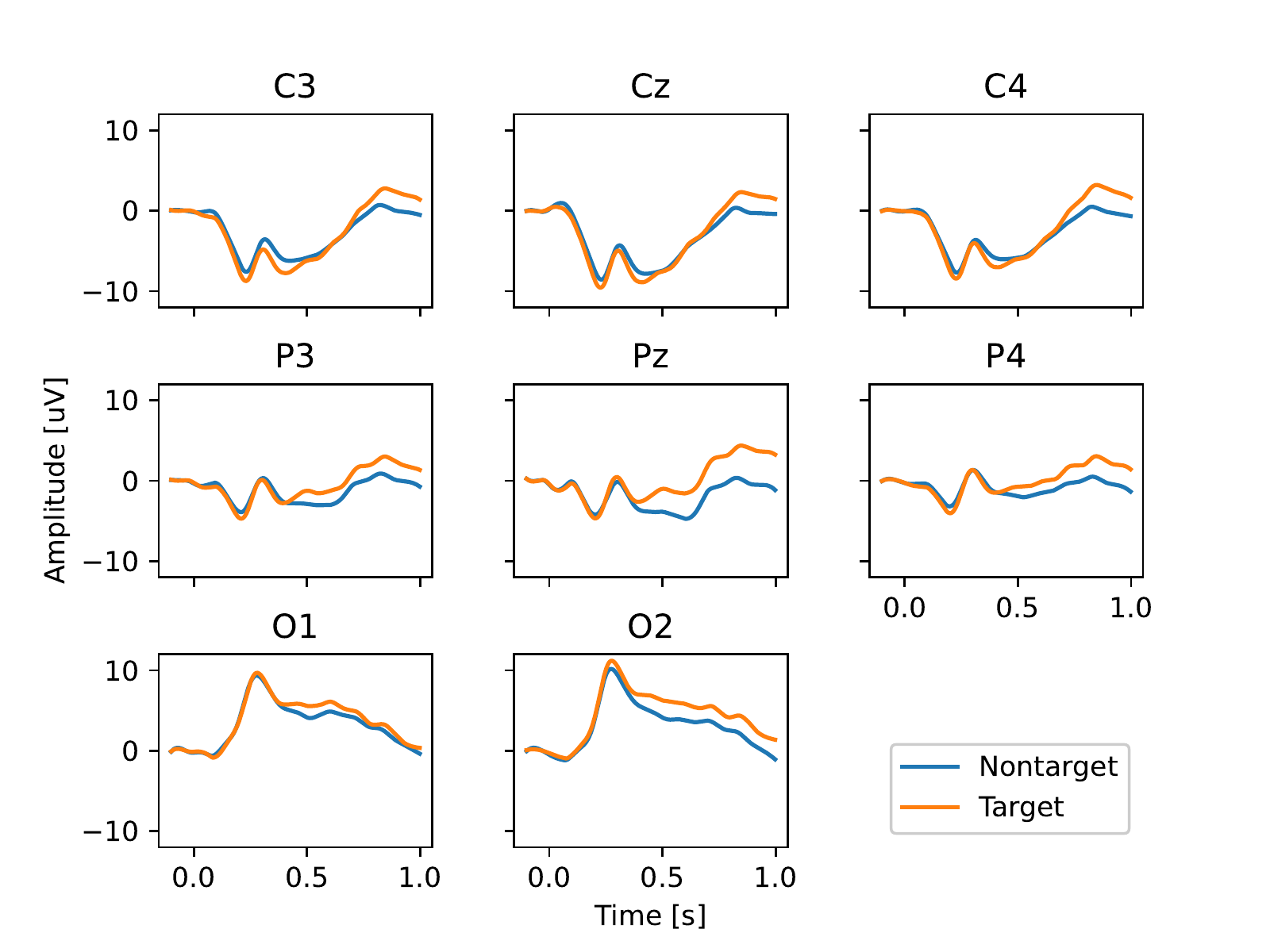}
	\caption{Grand-averaged EEG waveforms in the A condition. }
	\label{FIG:Exp1_GrandAverageA}
\end{figure}

\noindent
Grand-averaged EEG waveforms for AV, V, and A stimuli are 
shown in Fig.~\ref{FIG:Exp1_GrandAverageAV}, \ref{FIG:Exp1_GrandAverageV}, 
and \ref{FIG:Exp1_GrandAverageA}, respectively. 
In the AV and V condition, target and nontarget waveform differed between 0.4-0.6 s for most channels. 
P300 can be seen at the P3, Pz, and P4 in AV and V condition; however, no major differences were confirmed. 
EEG waveforms of V condition showed a smaller difference between target and nontarget, 
especially at C3, Cz, C4, and O1 channels.

\subsubsection{$r^2$-values}
\begin{figure}
	\centering
		\includegraphics[scale=.5]{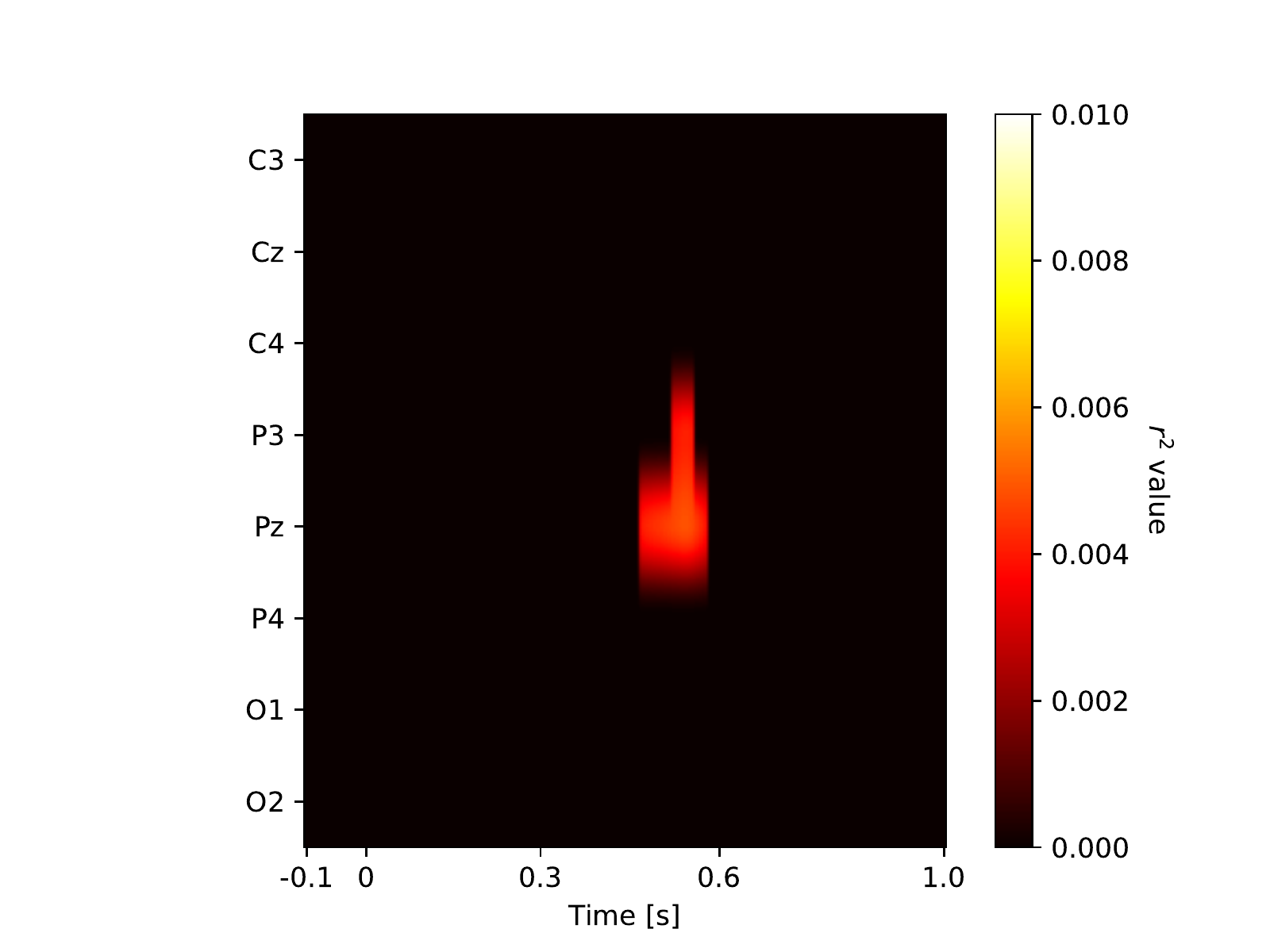}
	\caption{$r^2$-values in the AV condition. Only significant values are shown in color (above zero). 
	}
	\label{FIG:Exp1_BCC(AV)}
\end{figure}

\begin{figure}
	\centering
		\includegraphics[scale=.5]{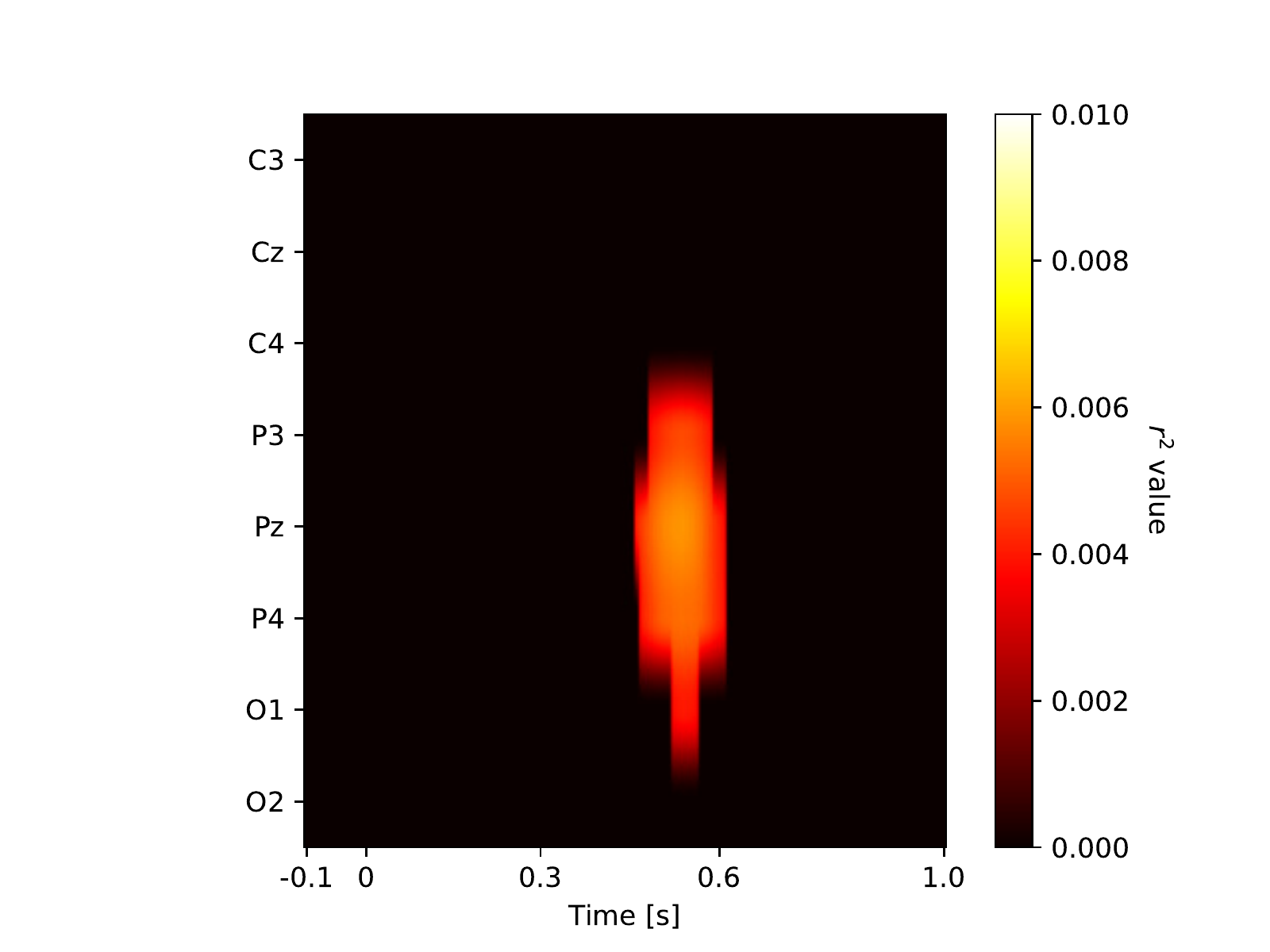}
	\caption{$r^2$-values in the V condition. Only significant values are shown in color (above zero). 
	}
	\label{FIG:Exp1_BCC(V)}
\end{figure}

\begin{figure}
	\centering
		\includegraphics[scale=.5]{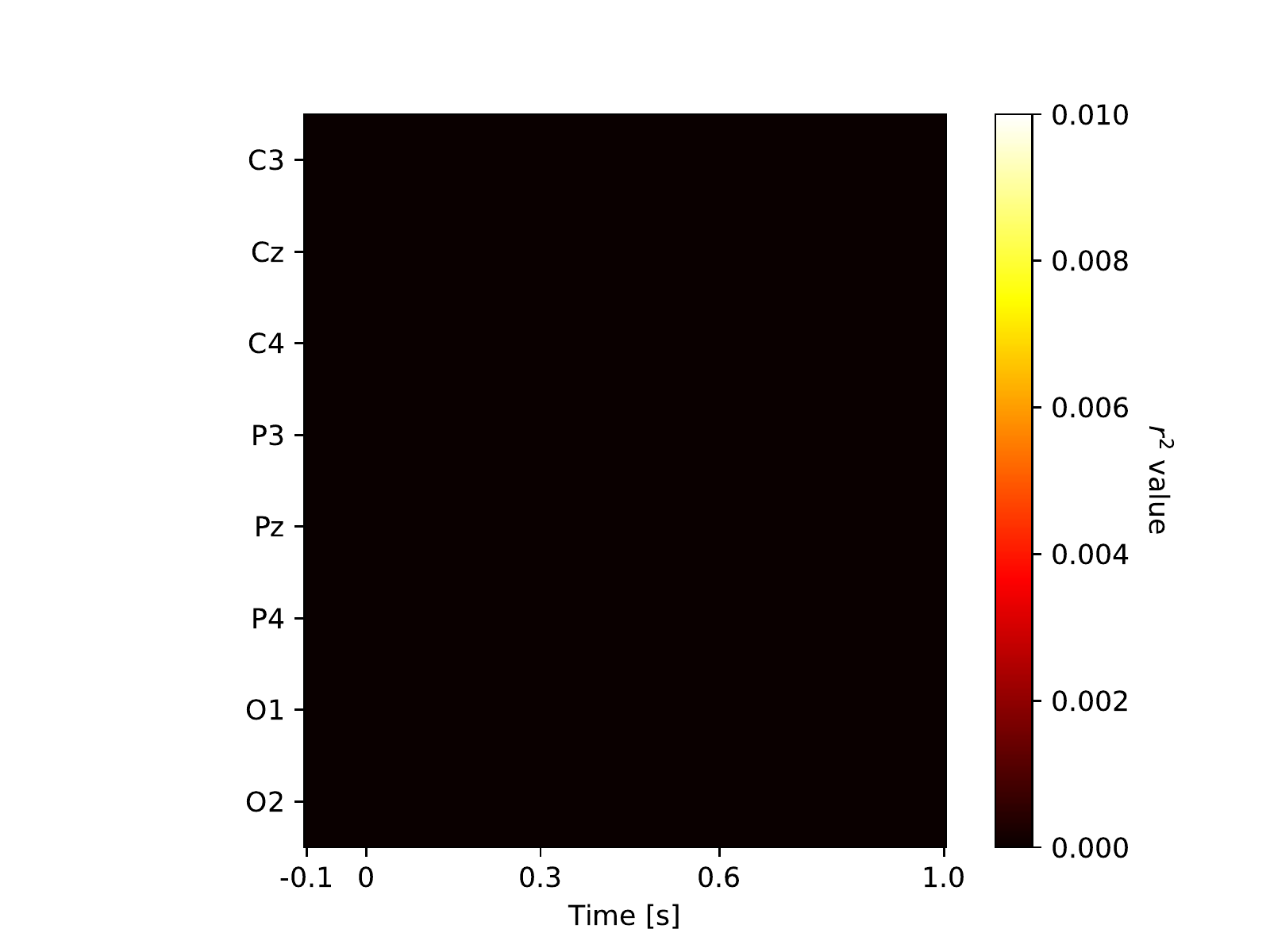}
	\caption{$r^2$-values in the A condition. Only significant values are shown in color (above zero). 
	}
	\label{FIG:Exp1_BCC(A)}
\end{figure}

\noindent
The biserial correlation coefficients for each stimulus condition
are represented in Fig.~\ref{FIG:Exp1_BCC(AV)}, \ref{FIG:Exp1_BCC(V)}, 
and \ref{FIG:Exp1_BCC(A)}, respectively. 
Significant $r^2$ values are shown in bright color, while non-significant $r^2$ values are shown in black (zero). 
In the AV condition, 
significant $r^2$ values were observed around 0.5 s for P3 and Pz. 
In addition, 
significant $r^2$ values appeared around 0.5 s for P3, Pz, P3, and Oz in the V condition. 
In contrast, no significant $r^2$ values were confirmed in the A condition.

\subsection{Experiment 2}

\subsubsection{Online classification accuracy} 
\begin{figure}
	\centering
		\includegraphics[scale=.5]{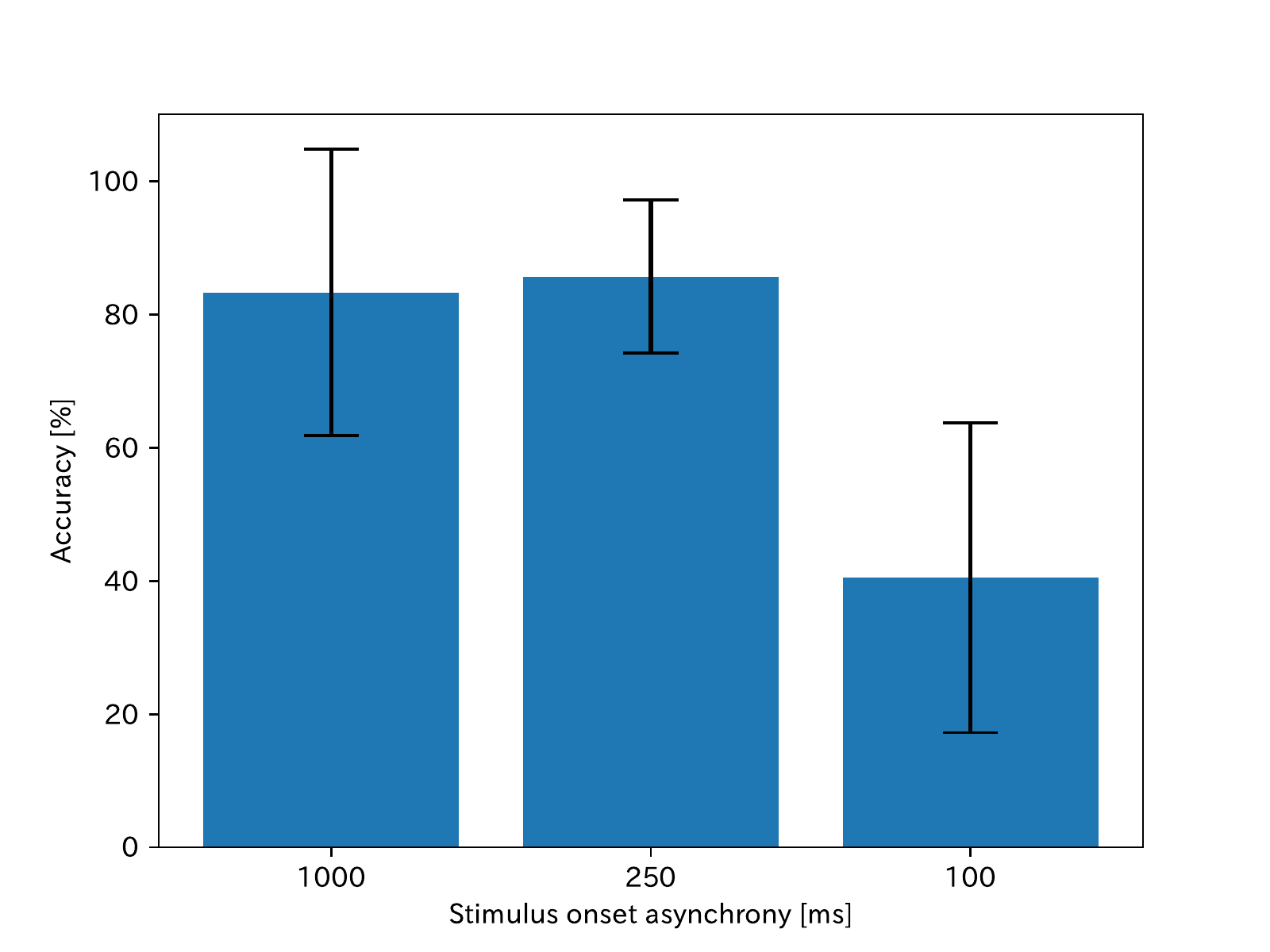}
	\caption{Online mean classification accuracy for SOA = 1000, 250, and 100 ms conditions.}
	\label{FIG:Exp2_Accuracy_Online}
\end{figure}

\noindent
Online classification accuracy was measured, changing SOA from 100 to 1000 ms. 
Figure \ref{FIG:Exp2_Accuracy_Online} and Table \ref{Table:Exp2_Online_acccuracy_SOA} represents the 
online classification accuracy when SOA is 1000, 250, and 100 ms. 
Some subjects achieved 100\% accuracy when SOA is 1000 and 250 ms. 
The mean classification accuracy was 83.3\%, 85.7\%, and 40.5\% 
for SOA of 1000, 250, and 100, respectively. 
The standard deviation was large when SOA was 1000, and 100.

This BCI system with SOA = 1000, 250 showed significantly higher accuracy than that with SOA = 100 ms. 
One-way repeated measures ANOVA indicated significant main effect of types of SOA ($p<0.001$, $F(2,12)$ = 16.36). 
The post-hoc pairwise t-test revealed significant differences between SOA = 100 and SOA = 250, and between SOA = 100 and SOA = 1000 ($p<0.05$).

\begin{table}[]
\caption{Online classification accuracy and standard deviation (SD) when SOA is 1000, 250, and 100 ms.}
\label{Table:Exp2_Online_acccuracy_SOA}
\begin{tabular*}{50mm}{C|CCC}
\toprule
Subject & \multicolumn{3}{c}{SOA [ms]} \\
 ~ & 1000 	& 250 	&100 \\
\hline
 1 	& 66.7	& 83.3 	& 16.7 \\
 2 	& 100.0	& 66.7	& 66.7 \\
 3 	& 100.0	& 100.0	& 66.7 \\
 4 	& 50.0	& 100.0	& 16.7 \\
 5 	& 100.0	& 83.3	& 50.0 \\
 6 	& 66.7	& 83.3	& 16.7 \\
 7 	& 100.0	& 83.3	& 50.0 \\
 Mean & 83.3	& 85.7	& 40.5 \\
 SD & 21.5	& 11.5	& 23.3 \\
\bottomrule
\end{tabular*}
\end{table}

\subsubsection{Online ITR}
\begin{figure}
	\centering
		\includegraphics[scale=.5]{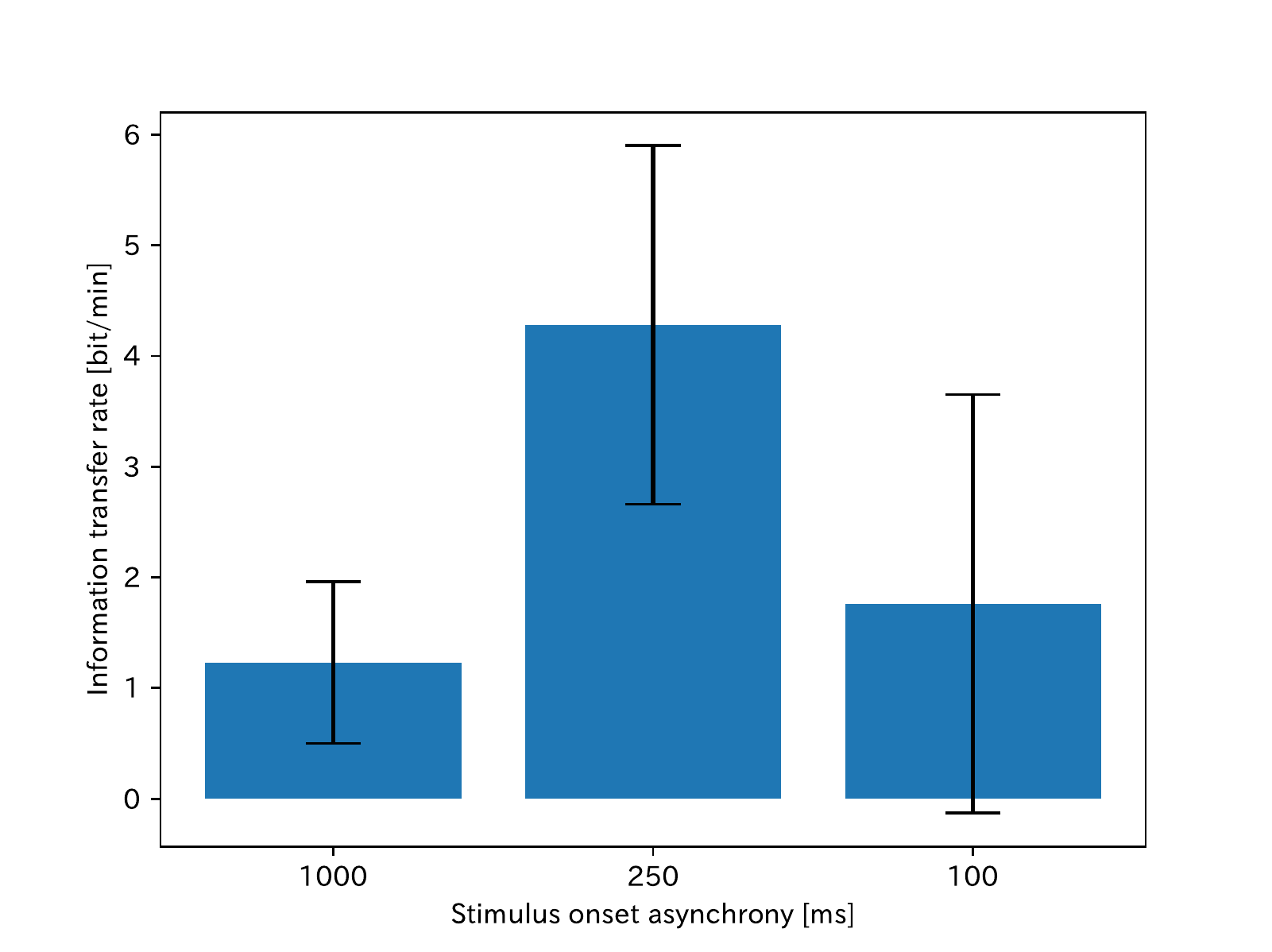}
	\caption{Online ITR for SOA = 1000, 250, and 100 ms conditions.}
	\label{FIG:Exp2_ITR_Online}
\end{figure}

\noindent
The online ITR was indicated in Fig.~\ref{FIG:Exp2_ITR_Online}. 
The mean online ITR was $1.23\pm0.73$, $4.28\pm1.6$, and $1.76\pm1.89$ bit/min  
when SOA was 1000, 250, and 100 ms, respectively. 

The ITR was highest when SOA was 250 ms. 
One-way repeated measures ANOVA indicated significant main effect of types of SOA ($p<0.001$, $F(2,12)$ = 8.523). 
The post-hoc pairwise t-test revealed significant differences between SOA = 100 and SOA = 250, and between SOA = 250 and SOA = 1000 ($p<0.05$).

\subsubsection{Offline classification accuracy} 
\begin{figure}
	\centering
		\includegraphics[scale=.5]{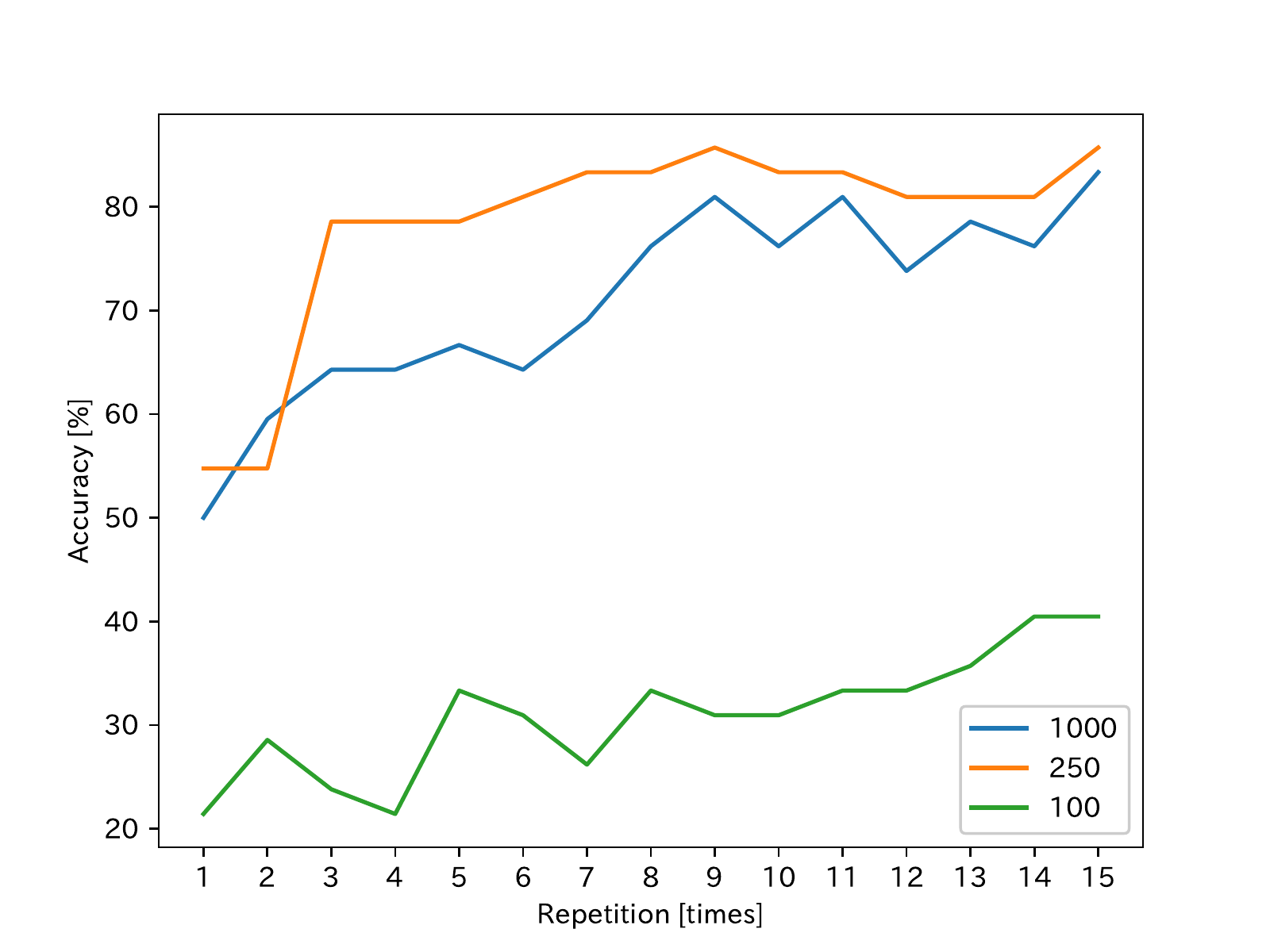}
	\caption{Offline mean classification accuracy for audiovisual stimuli when SOA = 1000, 250, and 100 ms.}
	\label{FIG:Exp2_Accuracy_Offfline}
\end{figure}

\noindent
EEG data recorded in exp.~2 was also analyzed offline by changing $R$ from 1 to 15. 
The results were shown in Table \ref{Table:Exp2_offline_acccuracy_SOA} and Fig.~\ref{FIG:Exp2_Accuracy_Offfline}. 
The offline accuracy with $R=15$ was equivalent to online classification accuracy indicated in Table \ref{Table:Exp2_Online_acccuracy_SOA}.
When SOA = 250 and $R=3$, mean accuracy reached 78.6\%. 
On the other hand, the BCI system with SOA = 1000 and 100 ms requires larger $R$ than that with $R=250$.   

The classification accuracy was the highest when SOA = 250. 
Two-way repeated measures ANOVA revealed significant main effects 
of stimulus type ($p<0.001$, $F(2,12)$ = 15.59) and repetition ($p<0.001$, $F(14,84)$ = 6.796). 
The post-hoc pairwise t-test revealed significant differences between all pairs of stimulus type ($p<0.05$).

\begin{table*}[width=2.0\linewidth,cols=17,pos=h]
\caption{Offline classification accuracy and standard deviation (SD) for each SOA and the number of repetition.}
\label{Table:Exp2_offline_acccuracy_SOA}
\begin{tabular*}{\tblwidth}{@{} CC|CCCCCCCCCCCCCCC@{} }
\toprule
SOA[ms] & Subject & \multicolumn{15}{c}{Repetition} \\
~ & ~ & 1 & 2 & 3 & 4 & 5 & 6 & 7 & 8 & 9 & 10 & 11 & 12 & 13 & 14 & 15 \\
\hline
1000 	& 1 & 50.0 & 66.7 & 50.0 & 50.0 & 50.0 & 50.0 & 66.7 & 66.7 & 66.7 & 66.7 & 66.7 & 66.7 & 66.7 & 66.7 & 66.7\\
~		& 2 	& 50.0 & 66.7 & 100.0 & 100.0 & 83.3 & 83.3 & 83.3 & 83.3 & 83.3 & 83.3 & 100.0 & 83.3 & 83.3 & 100.0 & 100.0\\
~		& 3 	& 83.3 & 83.3 & 100.0 & 100.0 & 100.0 & 100.0 & 100.0 & 100.0 & 100.0 & 100.0 & 100.0 & 100.0 & 100.0 & 100.0 & 100.0\\
~		& 4 	& 0.0 & 0.0 & 16.7 & 16.7 & 16.7 & 33.3 & 33.3 & 33.3 & 50.0 & 33.3 & 33.3 & 33.3 & 50.0 & 33.3 & 50.0\\
~		& 5 	& 83.3 & 83.3 & 66.7 & 83.3 & 100.0 & 100.0 & 100.0 & 100.0 & 100.0 & 100.0 & 100.0 & 100.0 & 100.0 & 100.0 & 100.0\\
~		& 6 	& 0.0 & 50.0 & 66.7 & 50.0 & 66.7 & 33.3 & 50.0 & 66.7 & 66.7 & 66.7 & 66.7 & 50.0 & 66.7 & 50.0 & 66.7\\
~		& 7 	& 83.3 & 66.7 & 50.0 & 50.0 & 50.0 & 50.0 & 50.0 & 83.3 & 100.0 & 83.3 & 100.0 & 83.3 & 83.3 & 83.3 & 100.0\\
~		& Mean & 50.0 & 59.5 & 64.3 & 64.3 & 66.7 & 64.3 & 69.0 & 76.2 & 81.0 & 76.2 & 81.0 & 73.8 & 78.6 & 76.2 & 83.3\\
~		& SD & 37.3 & 28.6 & 29.5 & 31.1 & 30.4 & 29.5 & 26.2 & 23.3 & 20.2 & 23.3 & 26.2 & 25.2 & 18.5 & 27.0 & 21.5\\
\hline
250 		& 1 & 66.7 & 50.0 & 66.7 & 66.7 & 66.7 & 66.7 & 83.3 & 100.0 & 66.7 & 66.7 & 83.3 & 66.7 & 66.7 & 83.3 & 83.3\\
~		& 2 	& 50.0 & 66.7 & 66.7 & 83.3 & 83.3 & 83.3 & 83.3 & 66.7 & 83.3 & 83.3 & 66.7 & 66.7 & 66.7 & 66.7 & 66.7\\
~		& 3 	& 66.7 & 50.0 & 83.3 & 83.3 & 83.3 & 83.3 & 100.0 & 100.0 & 100.0 & 100.0 & 83.3 & 83.3 & 83.3 & 83.3 & 100.0\\
~		& 4 	& 33.3 & 66.7 & 66.7 & 83.3 & 83.3 & 100.0 & 100.0 & 100.0 & 100.0 & 100.0 & 100.0 & 100.0 & 100.0 & 100.0 & 100.0\\
~		& 5 	& 66.7 & 50.0 & 83.3 & 66.7 & 66.7 & 66.7 & 66.7 & 66.7 & 83.3 & 66.7 & 83.3 & 83.3 & 83.3 & 66.7 & 83.3\\
~		& 6 	& 50.0 & 66.7 & 100.0 & 83.3 & 100.0 & 100.0 & 83.3 & 83.3 & 83.3 & 83.3 & 83.3 & 83.3 & 83.3 & 83.3 & 83.3\\
~		& 7 	& 50.0 & 33.3 & 83.3 & 83.3 & 66.7 & 66.7 & 66.7 & 66.7 & 83.3 & 83.3 & 83.3 & 83.3 & 83.3 & 83.3 & 83.3\\
~		& Mean & 54.8 & 54.8 & 78.6 & 78.6 & 78.6 & 81.0 & 83.3 & 83.3 & 85.7 & 83.3 & 83.3 & 81.0 & 81.0 & 81.0 & 85.7\\
~		& SD & 12.6 & 12.6 & 12.6 & 8.1 & 12.6 & 15.0 & 13.6 & 16.7 & 11.5 & 13.6 & 9.6 & 11.5 & 11.5 & 11.5 & 11.5\\
\hline
100 		& 1 & 0.0 & 0.0 & 33.3 & 16.7 & 16.7 & 16.7 & 16.7 & 16.7 & 16.7 & 16.7 & 33.3 & 16.7 & 16.7 & 16.7 & 16.7\\
~		& 2 	& 33.3 & 33.3 & 16.7 & 33.3 & 50.0 & 66.7 & 83.3 & 83.3 & 66.7 & 83.3 & 66.7 & 66.7 & 66.7 & 66.7 & 66.7\\
~		& 3 	& 16.7 & 0.0 & 0.0 & 16.7 & 16.7 & 0.0 & 0.0 & 33.3 & 33.3 & 33.3 & 16.7 & 16.7 & 16.7 & 50.0 & 66.7\\
~		& 4 	& 33.3 & 33.3 & 16.7 & 0.0 & 66.7 & 50.0 & 16.7 & 33.3 & 16.7 & 16.7 & 16.7 & 16.7 & 16.7 & 16.7 & 16.7\\
~		& 5 	& 33.3 & 50.0 & 33.3 & 33.3 & 33.3 & 33.3 & 33.3 & 16.7 & 33.3 & 16.7 & 33.3 & 33.3 & 50.0 & 50.0 & 50.0\\
~		& 6 	& 16.7 & 66.7 & 50.0 & 16.7 & 16.7 & 16.7 & 16.7 & 16.7 & 16.7 & 16.7 & 16.7 & 16.7 & 16.7 & 33.3 & 16.7\\
~		& 7 	& 16.7 & 16.7 & 16.7 & 33.3 & 33.3 & 33.3 & 16.7 & 33.3 & 33.3 & 33.3 & 50.0 & 66.7 & 66.7 & 50.0 & 50.0\\
~		& Mean & 21.4 & 28.6 & 23.8 & 21.4 & 33.3 & 31.0 & 26.2 & 33.3 & 31.0 & 31.0 & 33.3 & 33.3 & 35.7 & 40.5 & 40.5\\
~		& SD & 12.6 & 24.9 & 16.3 & 12.6 & 19.2 & 22.4 & 27.0 & 23.6 & 17.8 & 24.4 & 19.2 & 23.6 & 24.4 & 18.9 & 23.3\\
\bottomrule
\end{tabular*}
\end{table*}

\subsubsection{Offline ITR}
\begin{figure}
	\centering
		\includegraphics[scale=.5]{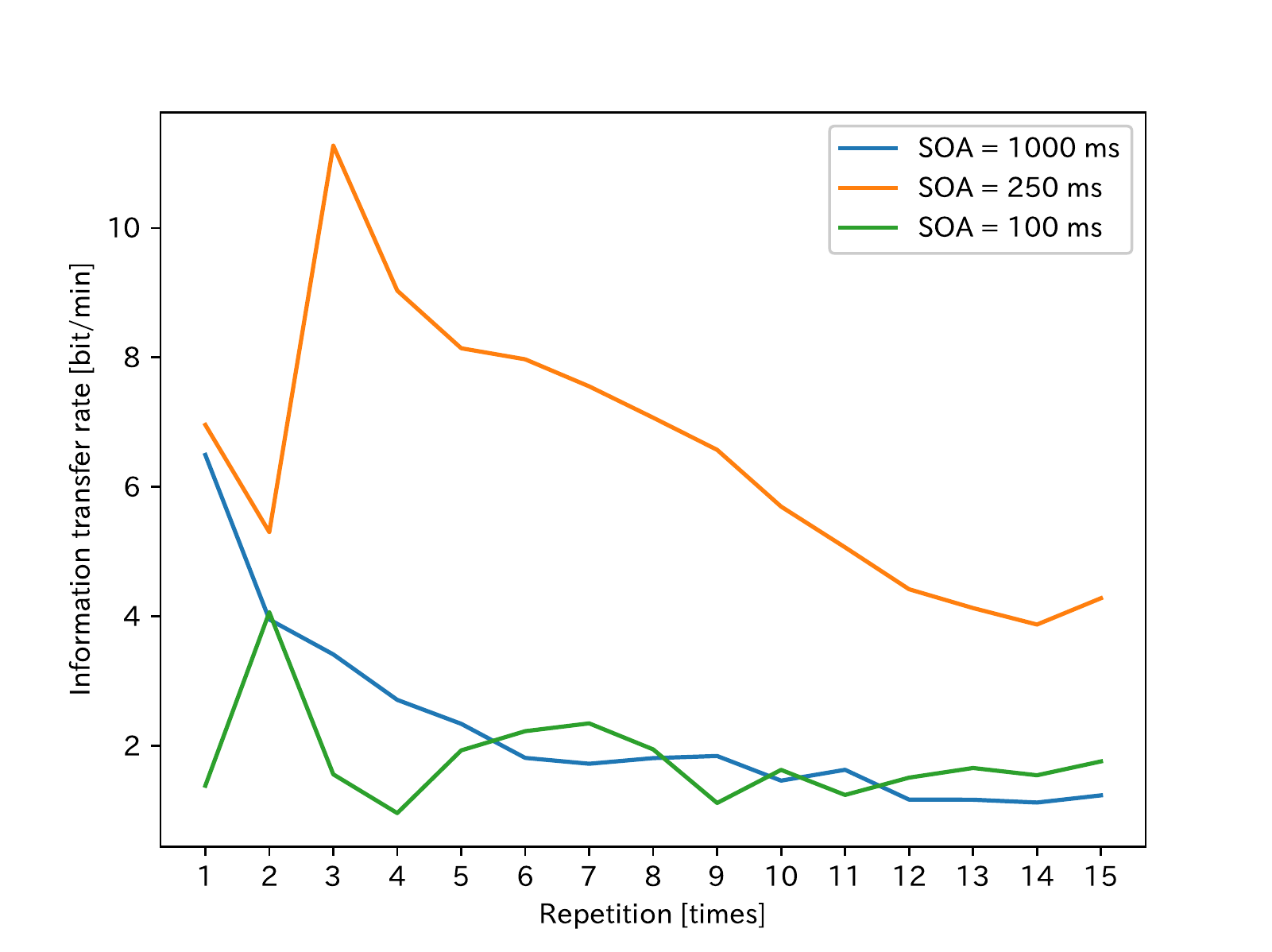}
	\caption{Offline mean ITR for audiovisual stimuli when SOA = 1000, 250, and 100 ms.}
	\label{FIG:Exp2_ITR_Offline}
\end{figure}

\noindent 
The mean offline ITR of exp.~2 was shown in Fig.~\ref{FIG:Exp2_ITR_Offline}. 
When SOA was 250 ms and the repetition was 3, the BCI achieved the highest mean ITR (11.3 bit/min). 

The BCI with SOA = 250 showed the highest ITR of all. 
Two-way repeated measures ANOVA revealed significant main effects 
of stimulus type ($p<0.001$, $F(2,12)$ = 18.14), repetition ($p<0.001$, $F(14,84)$ = 3.854), and their interaction ($p<0.001$, $F(28,168)$ = 2.719). 
The post-hoc pairwise t-test revealed significant differences between SOA = 100 and SOA = 250, and between SOA = 250 and SOA = 1000 ($p<0.05$). 

\subsubsection{Grand averaged EEG waveforms}
\begin{figure}
	\centering
		\includegraphics[scale=.5]{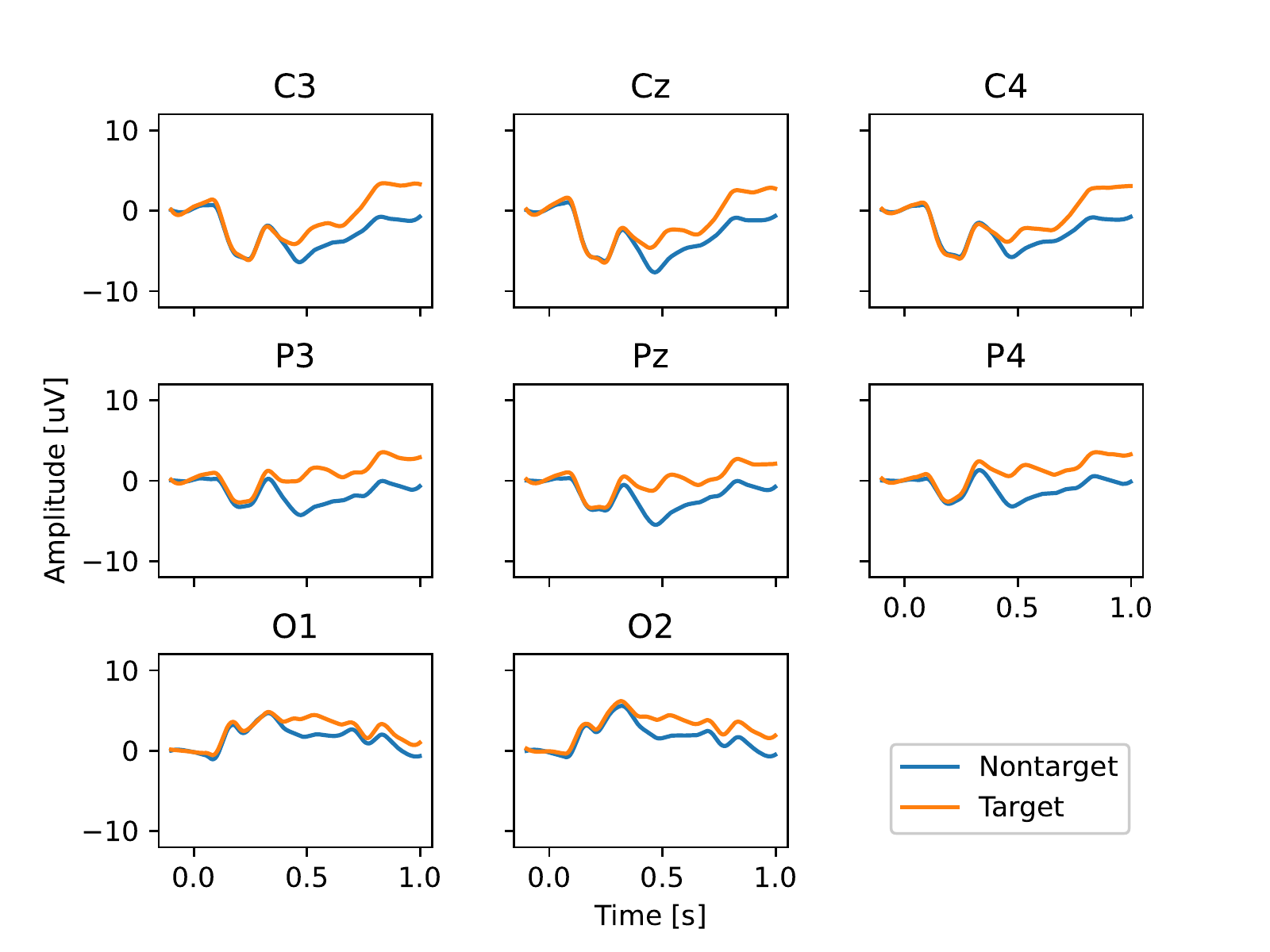}
	\caption{Grand-averaged EEG waveforms when SOA = 1000 ms. }
	\label{FIG:Exp2_GrandAverageAV1000}
\end{figure}

\begin{figure}
	\centering
		\includegraphics[scale=.5]{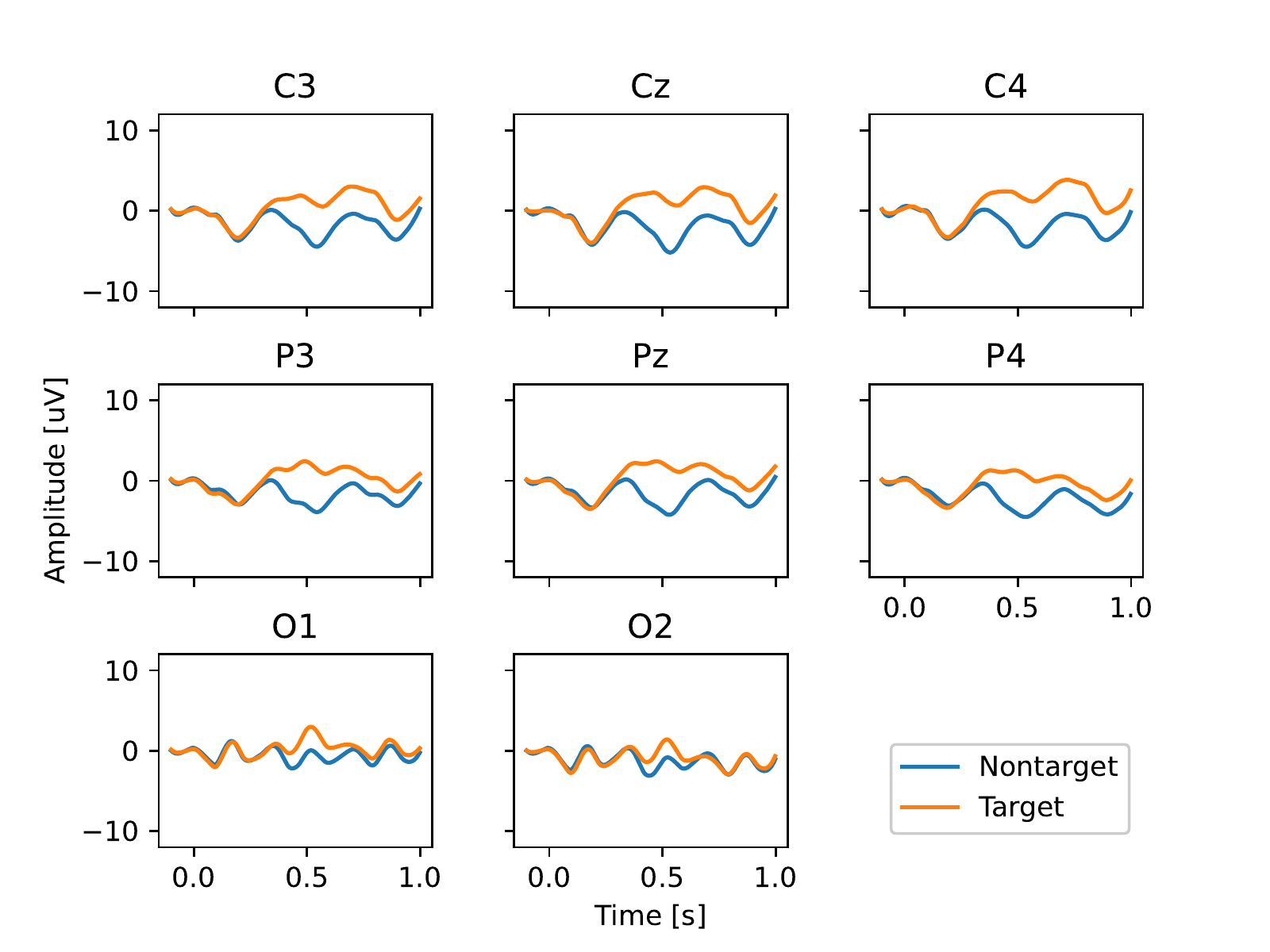}
	\caption{Grand-averaged EEG waveforms when SOA = 250 ms. }
	\label{FIG:Exp2_GrandAverageAV250}
\end{figure}

\begin{figure}
	\centering
		\includegraphics[scale=.5]{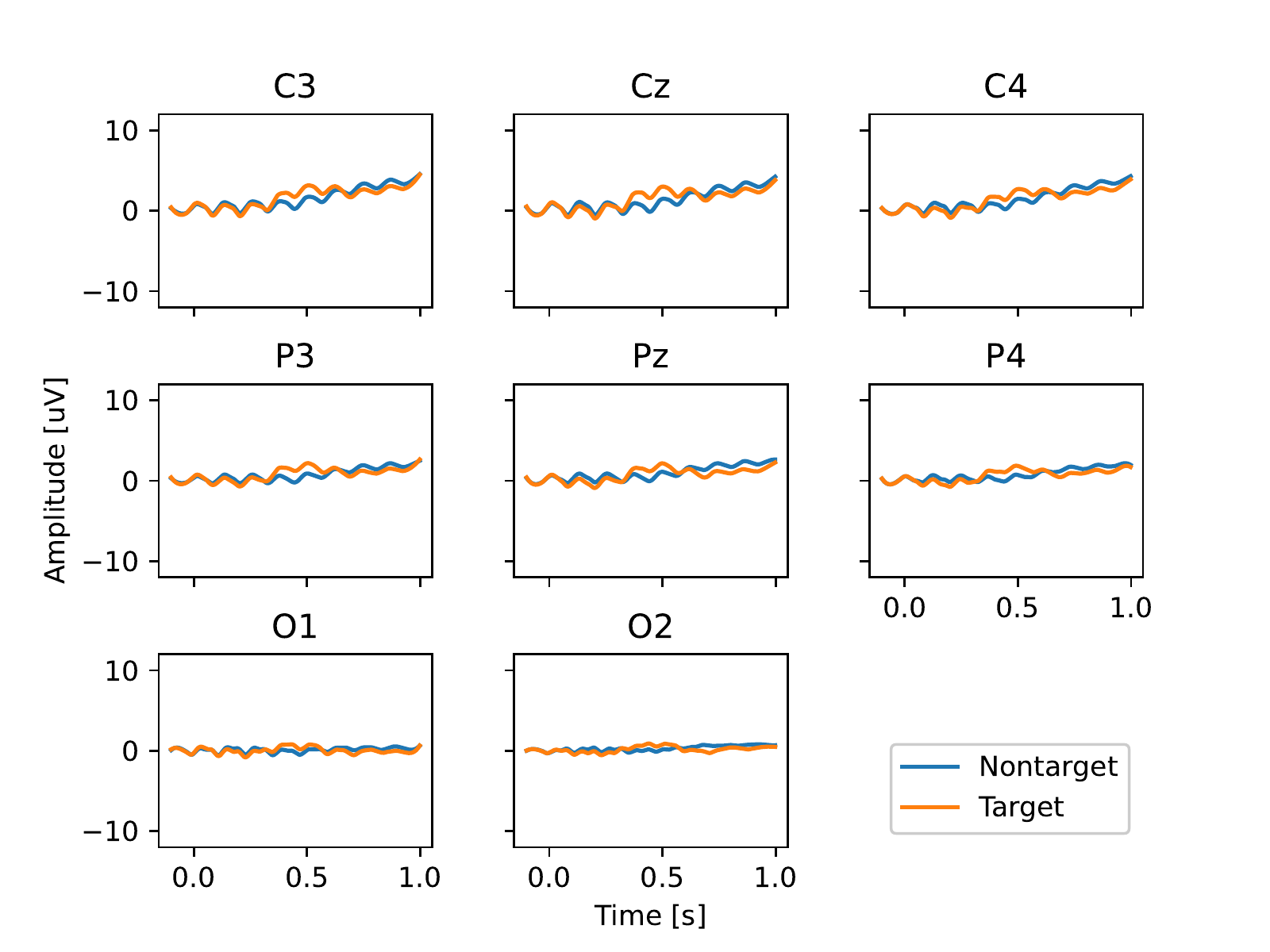}
	\caption{Grand-averaged EEG waveforms when SOA = 100 ms. }
	\label{FIG:Exp2_GrandAverageAV100}
\end{figure}

\noindent 
Grand averaged EEG waveforms of exp.~2 were shown in Fig.~\ref{FIG:Exp2_GrandAverageAV1000}, 
\ref{FIG:Exp2_GrandAverageAV250}, and \ref{FIG:Exp2_GrandAverageAV100}.
Target and nontarget differences were the largest when SOA was 250 ms. 
This was obvious between 0.4-0.6 s at C3, Cz, C4, P3, Pz, and P4. 
P300 was enhanced when SOA was 250 ms. 
When SOA = 1000 ms, target and nontarget differed between 0.4-1 s at C3, Cz, C4, P3, Pz, and P4. 
The difference was the smallest when SOA was 100 ms.

\subsubsection{$r^2$-values}
\begin{figure}
	\centering
		\includegraphics[scale=.5]{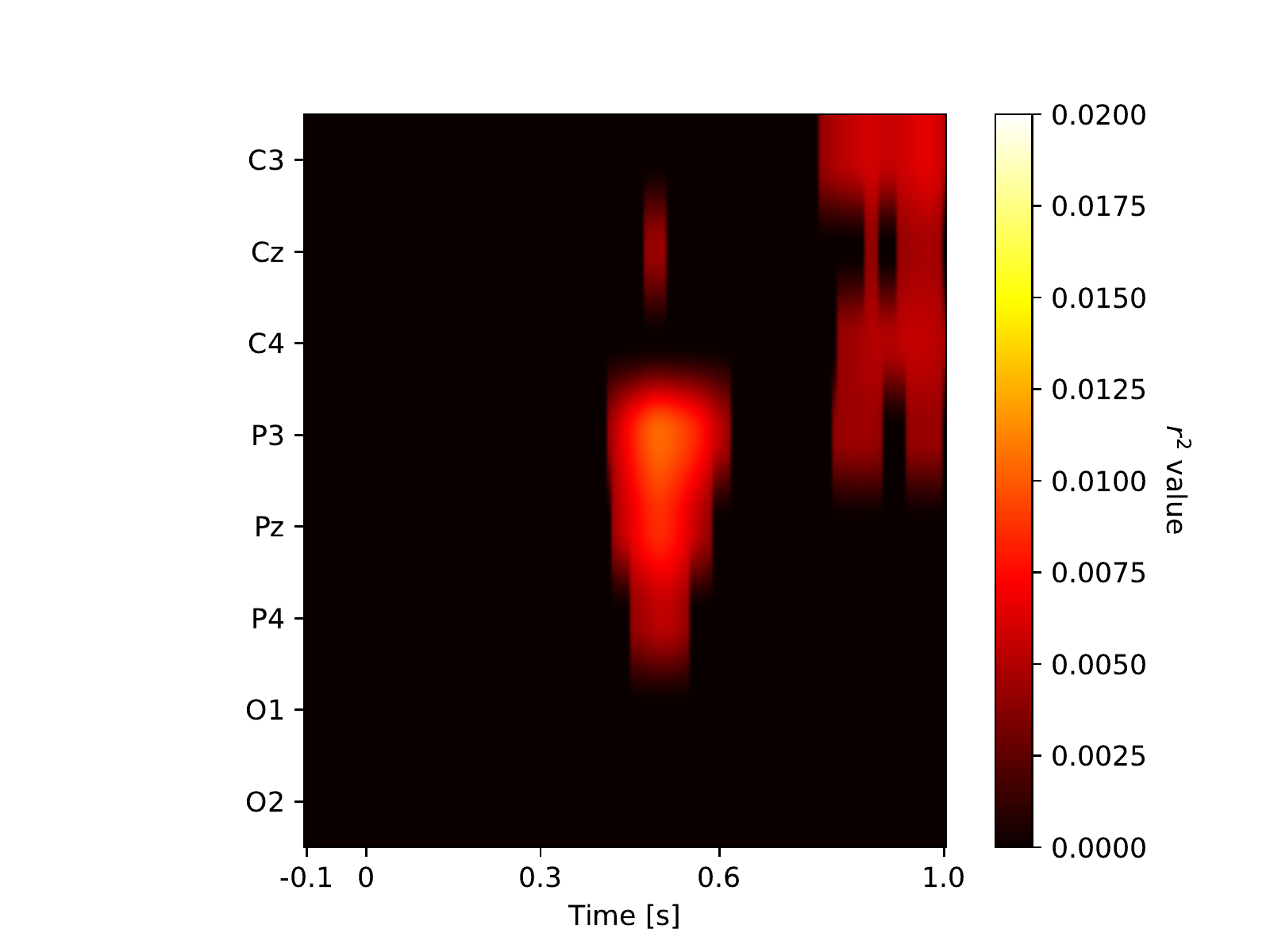}
	\caption{$r^2$-values when SOA = 1000 ms. Only significant values are shown in color (above zero). 
	}
	\label{FIG:Exp2_BCC(AV1000)}
\end{figure}

\begin{figure}
	\centering
		\includegraphics[scale=.5]{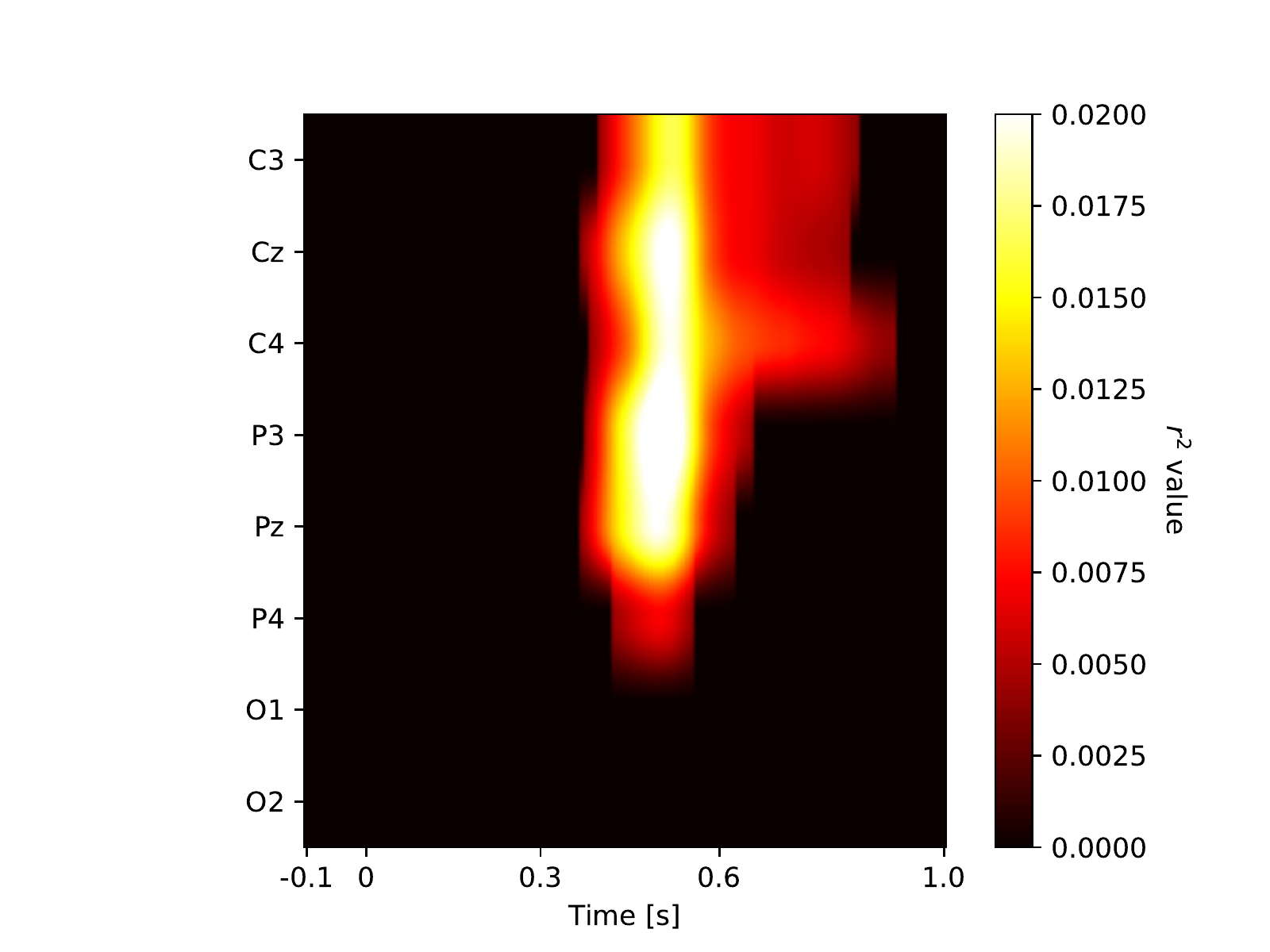}
	\caption{$r^2$-values when SOA = 250 ms. Only significant values are shown in color (above zero). 
	}
	\label{FIG:Exp2_BCC(AV250)}
\end{figure}

\begin{figure}
	\centering
		\includegraphics[scale=.5]{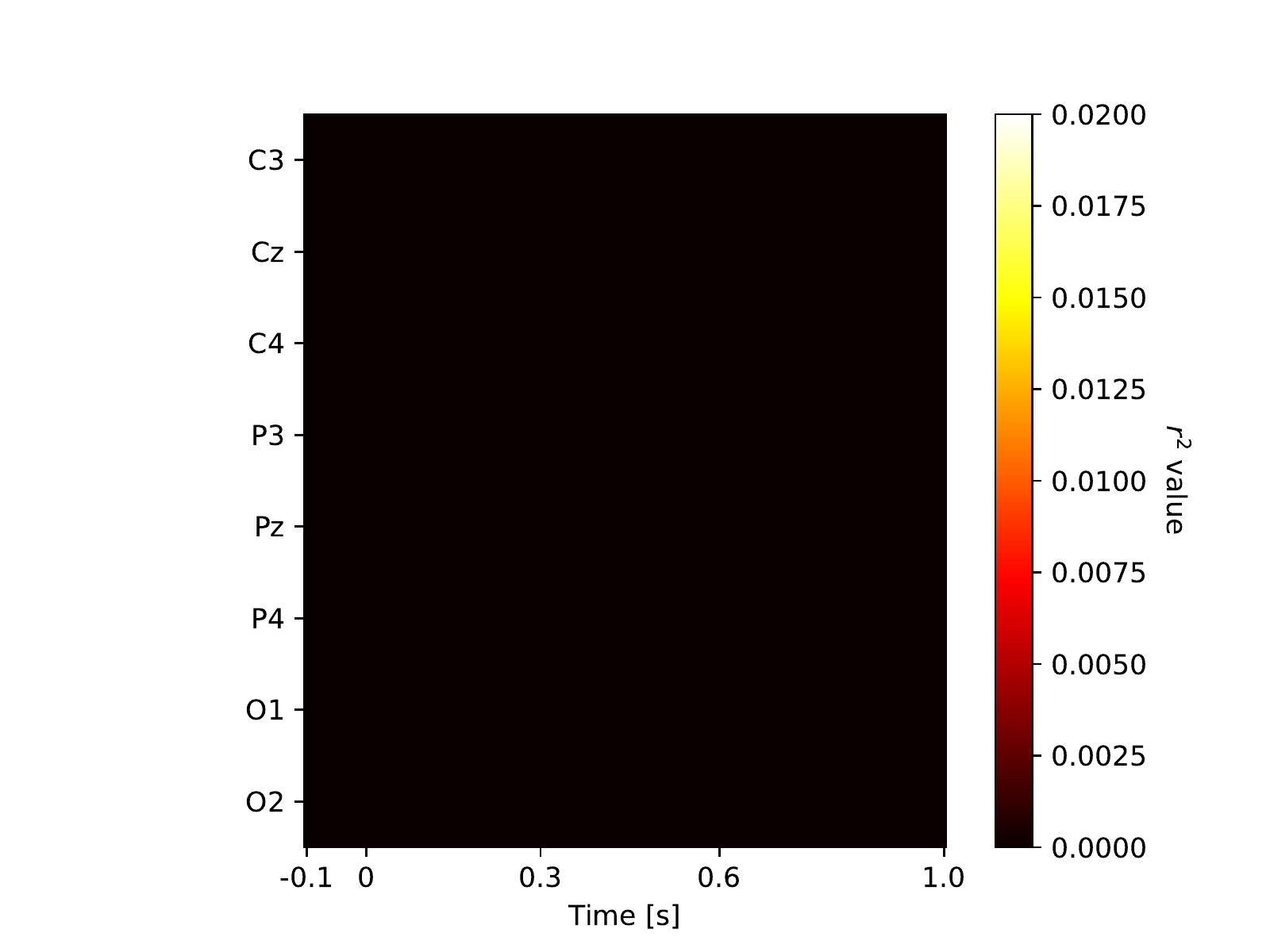}
	\caption{$r^2$-values when SOA = 100 ms. Only significant values are shown in color (above zero). 
	}
	\label{FIG:Exp2_BCC(AV100)}
\end{figure}

\noindent 
The biserial correlation coefficients of exp.~2 
are represented in Fig.~\ref{FIG:Exp2_BCC(AV1000)}, \ref{FIG:Exp2_BCC(AV250)}, and \ref{FIG:Exp2_BCC(AV100)}, respectively. 
The bright color of those figures indicated significant $r^2$ values. 
Note that non-significant $r^2$ values are drawn in black. 
When SOA was 250 ms, significant $r^2$ values were confirmed in a broader area between 0.4-0.6 s at C3, Cz, C4, P3, Pz, and P4. 
When SOA was 1000 ms,  significant $r^2$ values appeared between 0.4-0.6 s at P3, Pz, and P4, and between 0.8-1.0 s at C3, Cz, and C4. 
No significant $r^2$ values were confirmed when SOA was 100 ms.

\section{Discussion}
\noindent
In the present study, we investigated the integrated effect of audiovisual content in an RSMP BCI system.
Our findings indicated that the highest classification accuracy occurred in the AV condition.
This result implies that sensory integration of auditory and visual contents increases the classification accuracy, even for RSMP BCIs. 
The online classification accuracy was 85.7$\pm$11.5\% when SOA was 250 ms. 
The offline mean ITR was 11.3 bit/min at best (SOA = 250 ms, $R$=3).  
Since RSMP can be used in gaze-independent BCIs, our study provides insight into methods that can be used to develop a new gaze-independent BCIs. 

Our findings indicated that the AV condition was associated with the best performance among all three conditions, 
in accordance with the findings of previous P300-based BCI studies.
Wang et al.~reported that an audiovisual P300-based BCI exhibited better performance 
than a visual-only or auditory-only BCI \cite{Wang2015}. 
However, this may have been due to the absence of visual or auditory stimuli. 
Even though the current study included both visual and auditory stimuli, 
the AV condition yielded the greatest classification accuracy. 
Thurlings et al. examined a visual-tactile BCI in congruent and incongruent conditions, 
showing that the congruent condition yielded performance improvements \cite{Thurlings2014}. 
These results imply that an absence of stimuli or existing 
yet incomprehensible stimuli do not contribute to increases in BCI performance.
In other words, pairs of multimodal stimuli that can be integrated easily 
may help increase BCI accuracy. 

In AV and V condition of exp.~1, P300 was found around Pz with significant $r^2$ values. 
However, no major difference was confirmed between the two. 
A bimodal P300-based BCI incorporating both tactile and visual stimuli exhibited a P300 
in addition to slight enhancement of P300 when compared with unimodal BCI \cite{Brouwer2010}. 
An auditory-tactile BCI indicated enhancement of the P300 at Fz \cite{Jiang2019}.
A study reported that enhancement of the N1 component and reduction of P300 amplitude in a visual-tactile P300-based BCI \cite{Thurlings2012}. 
Exp.~1 of the current study did not indicate P300 enhancement, 
which was not in line with most multimodal studies.

The enhancement of P300 amplitude depends on SOA for RSMP-BCI. 
Results of exp.~2 indicated that the ERP waveform was influenced by the length of the SOA. 
Since ERPs elicited by a stimulus lasted approximately 1000 ms, the overlap of ERP caused by the SOA should change ERP shapes. 
As a result, exogenous ERP components remained in grand-averaged ERP waveforms when SOA was longer. 
The exogenous ERP component will look small 
if the SOA is short because the overlap of each stimulus is large. 
In exp.~2, P300 was enhanced only when SOA was 250 ms.
Some RSVP BCI studies have reported SOAs of only 100 ms \cite{Zheng2020, Wei2020}, which elicited P300. 
However, SOA = 100 ms was not a suitable parameter for the current BCI because the BCI had only 5 stimuli, and participants were required to count twice a second. 
In other words, the mental task of the current BCI was too difficult when SOA = 100. 
Researchers have examined the influence of speed on RSVP, 
reporting that the area under the curve decreases as SOA increases \cite{Lees2020}. 
However, no studies have investigated SOAs longer than 500 ms. 
The current BCI worked even when SOA was 1000 ms, showing P300 around Pz.

In the current study, we observed no face-specific ERP components, in contrast to previous studies reporting face specificity for N170 and N400 \cite{Jin2012}. This may be explained by (1) the effect of audiovisual stimuli, (2) the effect of artificial facial images, or (3) the effect of the manner of stimulus presentation. 
However, face-specific ERP components were observed 
when visual stimuli were presented without auditory stimuli, indicating that these components may be changed by auditory stimuli. 
Alternatively, the component may not appear when using artificial facial images.
One study reported that both dummy facial images 
and facial images could evoke similar ERP components \cite{chen2015survey}. 
Another study reported that an RSVP BMI using facial images exhibited N170 components \cite{cai2013rapid}. 
Therefore, the use of audiovisual stimuli may have resulted in the disappearance of face-specific ERP components.

\section{Conclusion}
\noindent
In the present study, we proposed an RSMP BCI 
that utilizes artificial facial images and artificial voice. 
To clarify the effect of audiovisual stimuli on the BCI, 
scrambled images and masked sounds were employed as stimuli, respectively. 
Our results indicated that audiovisual stimuli without scrambled images or masked sounds yielded the highest classification accuracy for the RSMP BCI. 
These results suggest the feasibility of audiovisual stimuli for use with RSMP BCIs and may help to improve gaze-independent BCI systems.

\printcredits

\bibliographystyle{cas-model2-names}

\bibliography{refs}

\end{document}